\documentclass[12pt]{article}
\usepackage{amsmath,amssymb,epsfig}
\makeatletter \@addtoreset{equation}{section} \makeatother
\renewcommand{\theequation}{\thesection.\arabic{equation}}
\newcommand{\ba}{\begin{array}}
\newcommand{\ea}{\end{array}}
\newcommand{\beq}{\begin{equation}}
\newcommand{\eeq}{\end{equation}}
\newcommand{\bea}{\begin{eqnarray}}
\newcommand{\eea}{\end{eqnarray}}

\def\bce{\begin{center}}
\def\ece{\end{center}}

\def\nonu{\nonumber}

\def\be{\beta}

\def\eps6{{\displaystyle \mathop{\epsilon}^{6}}{}}

\def\nab6{{\displaystyle \mathop{\nabla}^{6}}{}}

\def\0{{\sst{(0)}}}
\def\1{{\sst{(1)}}}
\def\2{{\sst{(2)}}}
\def\3{{\sst{(3)}}}
\def\4{{\sst{(4)}}}
\def\5{{\sst{(5)}}}
\def\6{{\sst{(6)}}}
\def\7{{\sst{(7)}}}
\def\8{{\sst{(8)}}}

\def\ba{\begin{array}}
\def\ea{\end{array}}
\def\beq{\begin{equation}}
\def\eeq{\end{equation}}
\def\be{\begin{equation}}
\def\ee{\end{equation}}

\def\eps{\epsilon}

\def\ba{\begin{array}}
\def\ea{\end{array}}
\def\beq{\begin{equation}}
\def\eeq{\end{equation}}
\def\be{\begin{equation}}
\def\ee{\end{equation}}

\def\eps{\epsilon}

\newcommand{\bean}{\begin{eqnarray*}}
\newcommand{\eean}{\end{eqnarray*}}

\begin{document}
\thispagestyle{empty} \addtocounter{page}{-1}
\begin{flushright}
\end{flushright}

\vspace*{1.3cm}

\centerline{ \Large \bf Meta-Stable Brane Configurations with 
Five NS5-Branes   }
\vspace*{1.5cm}
\centerline{{\bf Changhyun Ahn} 
} 
\vspace*{1.0cm} 
\centerline{\it 
Department of Physics, Kyungpook National University, Taegu
702-701, Korea} 
\vspace*{0.8cm} 
\centerline{\tt ahn@knu.ac.kr} 
\vskip2cm

\centerline{\bf Abstract}
\vspace*{0.5cm}

From an ${\cal N}=1$ supersymmetric electric gauge theory 
with the gauge group $SU(N_c) \times SU(N_c')$ with fundamentals for each 
gauge group, the bifundamentals and a symmetric flavor and a
conjugate
symmetric flavor for $SU(N_c)$, we apply Seiberg dual 
to  each gauge group  independently and obtain two ${\cal N}=1$
supersymmetric 
dual magnetic gauge theories with dual matters including the gauge
singlets. 
By analyzing the F-term equations of the dual
magnetic 
superpotentials, we describe the intersecting brane configurations of
type 
IIA string theory corresponding to the meta-stable nonsupersymmetric 
vacua of these gauge theories.
The case where the above 
symmetric flavor is replaced by an antisymmetric flavor is also discussed.  

\baselineskip=18pt
\newpage
\renewcommand{\theequation}
{\arabic{section}\mbox{.}\arabic{equation}}

\section{Introduction}

Starting from ${\cal N}=1$ supersymmetric gauge theory with massive
fundamentals, the construction of meta-stable supersymmetry breaking
vacua was found in \cite{ISS} by using the corresponding 
dual magnetic gauge theory. The magnetic theory does have
superpotential
consisting of an interaction between the meson and dual quarks as well as 
a linear term in the meson that can be interpreted as a mass term for
the quarks in the electric theory. 
By rank condition, the F-term equation from the dual magnetic
superpotential cannot be satisfied and the
supersymmetry is broken. See the review paper \cite{IS} for 
the recent developments of dynamical supersymmetry breaking.

In the type IIA brane configuration \cite{GK}, the above gauge theory can be
described by two NS5-branes, D6-branes and D4-branes.
By standard brane motion,
the magnetic brane configuration or Seiberg dual can be constructed 
from electric one.
The deformation of quark mass corresponds to the relative displacement
of  D6-branes and D4-branes along the common orthogonal directions.
The geometric misalignment of flavor D4-branes in the magnetic brane
configuration
can be interpreted as a
nontrivial F-term equation we mentioned in the magnetic gauge theory. 

When an adjoint matter is included, then 
one should consider a set of coincident NS5-branes rather than a
single NS5-brane. 
When we add an orientifold 4-plane(O4-plane) to the above brane configuration,
then the gauge group will be changed into a symplectic or orthogonal
gauge groups. On the other hand, if we include an orientifold 6-plane(O6-plane), 
then the matter contents will be different due to the projection. Totally, the three
NS5-branes are present and a middle NS5-brane is located at an
orientifold 6-plane for the unitary gauge group \cite{GK}.  
All of these considerations have a single gauge group with
corresponding matters.

What happens when we consider product of two gauge groups?
Without any orientifold plane, there exist three NS5-branes, D6-branes
and D4-branes. 
When we add an orientifold 4-plane to this brane configuration,
then the gauge group will be changed into product gauge group of 
a symplectic and orthogonal
gauge groups. On the other hand, if we include an orientifold 6-plane, 
then the matter contents will be different, in general.
When there is no NS5-brane on an orientifold 6-plane, four NS5-branes
are present and the gauge group will be a product of unitary gauge
group and orthogonal or symplectic gauge group. On the other hand, if
the NS5-brane is located at an orientifold 6-plane, then five
NS5-branes are needed and the gauge group will be a product  of
unitary gauge groups.
   
In this paper, we consider a particular product gauge group
$SU(N_c) \times SU(N_c')$ with 
 fundamentals for each 
gauge group, the bifundamentals as well as a symmetric  and conjugate
symmetric flavor for $SU(N_c)$.
Without these symmetric and conjugate symmetric flavors, the type IIA
brane configuration for this product gauge group with fundamentals
and bifundamentals consists of three NS5-branes, D6-branes and
D4-branes for each gauge group \cite{BH,GK}. For purely gauge theory
analysis, see \cite{BIWW,ILS} for details.
On the other hand, if we ignore the second gauge group $SU(N_c')$ with
corresponding matter contents completely, 
this theory will reduce to a single gauge group $SU(N_c)$ with a
symmetric 
flavor,
conjugate symmetric flavor and fundamental flavors developed in \cite{ILS,LLL}:
there are three NS5-branes, D6-branes, D4-branes and an
orientifold 6-plane where a middle NS5-brane is located.
In addition to these branes,
we add the extra two outer NS5-branes in a ${\bf Z}_2$ symmetric
way due to the O6-plane, extra D4-branes 
and extra D6-branes corresponding to the second gauge group $SU(N_c')$.
Starting from this ${\cal N}=1$ supersymmetric gauge theory with massive
fundamentals for the each gauge group, 
we will analyze the meta-stable supersymmetry breaking brane configuration.

The product gauge group
$SU(N_c) \times SU(N_c')$ with 
fundamentals for each 
gauge group, the bifundamentals as well as an antisymmetric  and conjugate
symmetric flavor for $SU(N_c)$ is also considered.

In section 2, we describe the type IIA brane configuration corresponding
to the electric theory based on the ${\cal N}=1$ $SU(N_c) \times
SU(N_c')$ 
gauge theory 
with matter contents and deform this theory by adding the mass term
for the quarks for each gauge group. 
Then we construct the Seiberg dual magnetic theory which is 
${\cal N}=1$ $SU(\widetilde{N}_c) \times SU(N_c')$ gauge 
theory with corresponding dual
matters as well as various gauge singlets, by brane motion and linking
number counting. 
Similarly, we construct the Seiberg dual magnetic theory which is 
${\cal N}=1$ $SU(N_c) \times SU(\widetilde{N}_c')$ gauge 
theory with corresponding dual
matters as well as various gauge singlets.

In section 3,
we consider the nonsupersymmetric meta-stable
minimum by looking at the magnetic brane configurations we obtained in
section 2, present 
the corresponding intersecting brane configurations of type IIA string
theory, and describe M-theory lift of this supersymmetry breaking 
type IIA brane configurations, along the line of 
\cite{Ahn07-3,Ahn07-2,Ahn07-1,Ahn07,Ahn06-1,Ahn06}.
The role of flavor D4-branes, i.e., a misalignment of these D4-branes, 
is crucial to describe these brane 
configurations. 

In section 4,  we summarize what we have done in previous sections.
We describe the similar application to the 
same ${\cal N}=1$ $SU(N_c) \times SU(N_c')$ 
gauge theory  with different matter contents, in the sense that the
above symmetric flavor is replaced by eight fundamentals and an
antisymmetric flavor for the $SU(N_c)$ gauge group. The theory 
given by the first gauge group $SU(N_c)$ with matters is based on 
the previous works of \cite{LLL1,BHKL,EGKT}.  
We also make some comments for the future directions.  

\section{The ${\cal N}=1$ supersymmetric  
brane configurations}

In order to study the meta-stable brane configuration, it is necessary
to take two steps. One of them is to have nonzero masses for the
quarks corresponding to relative displacement between D6-branes and
D4-branes
and the other is to take the Seiberg dual theory by standard brane
motion. 
For the
latter, we need to understand both brane configurations from the
electric theory and the magnetic theory since the magnetic theory can
be obtained from the electric theory \cite{GK}. These brane configurations 
corresponding to the gauge theory we are considering 
are not known so far in the literature and we describe them more explicitly. 
There exist two possible magnetic brane configurations, depending on
whether the dual gauge group we take is the first gauge group  or the second
gauge group. Note that although the gauge group is a product
gauge group, it is not always possible to take the dual for each of
the 
gauge group independently, in the context of meta-stable brane
configuration and for example, see \cite{Ahn07-3}.  

\subsection{Electric theory  with $SU(N_c) \times SU(N_c')$
  gauge group}

The gauge group we are interested in is given by $SU(N_c) \times
SU(N_c')$ and 
the matter contents
are as follows:

$\bullet$
$N_f$-chiral multiplets $Q$ are  in the
representation $({\bf N_c,1
})$, and 
$N_f$-chiral multiplets $\widetilde{Q}$ are in  
the representation $({\bf \overline{N_c}, 1})$,
under the gauge group

$\bullet$
$N_f'$-chiral multiplets $Q'$ are  in the
representation $({\bf 1, N_c'
})$, and 
$N_f'$-chiral multiplets $\widetilde{Q'}$ are in  
the representation $({\bf 1,\overline{N_c'}})$,
under the gauge group

$\bullet$
The flavor-singlet field $X$ is 
in the bifundamental representation $({\bf N_c, \overline{N_c'} })$, 
and its conjugate field $\widetilde{X}$
 is 
in the bifundamental representation $({\bf \overline{N_c}, N_c'})$, 
under the gauge group

$\bullet$ The flavor-singlet field $S$, which is 
in a symmetric tensor representation under the $SU(N_c)$, is in the
representation $({\bf \frac{1}{2} N_c(N_c+1),1})$, and
its conjugate field $\widetilde{S}$ is in the 
representation $({\bf \overline{\frac{1}{2} N_c(N_c+1)},1})$, under the
gauge group

If there are no symmetric and conjugate symmetric tensors, $S$ and
$\widetilde{S}$, 
this theory 
is described by the work of \cite{ILS,BH,BIWW} from field theory analysis or
corresponding brane
dynamics.  Ignoring the presence of the fields $Q', \widetilde{Q'}, X$ and
$\widetilde{X}$,
then this theory will reduce to a single gauge group $SU(N_c)$ with a symmetric flavor,
conjugate symmetric flavor and fundamental flavors $S, \widetilde{S},
Q$ and $\widetilde{Q}$ discussed in \cite{ILS,LLL,Ahn07}.
Now it is easy to check that 
the coefficient of the beta function of the first gauge group 
is given by
\bea
b_{SU(N_c)}=3N_c-N_f-N_c'-(N_c+2)
\nonu
\eea
where the index of the symmetric representation of 
$SU(N_c)$ corresponding to $S$ and $\widetilde{S}$ is equal
to $\frac{1}{2}(N_c+2)$.
On the other hand,
the coefficient of the beta function of the second gauge group  
is given by
\bea
b_{SU(N_c')}=3N_c'-N_f'-N_c.
\nonu
\eea
This theory is asymptotically free when the condition $b_{SU(N_c)} >
0$ is satisfied for the
$SU(N_c)$
gauge group and when the condition $b_{SU(N_c')} > 0$ is satisfied 
for the $SU(N_c')$ gauge group.
We'll see how these coefficients change in the magnetic theory.

The classical superpotential is given by
\bea
W= \mu A^2 + S A \widetilde{S} + \lambda Q A \widetilde{Q}
+ \mu' A^{'2} + \lambda' Q' A' \widetilde{Q'} + X A \widetilde{X} + 
\widetilde{X} A' X
+ m Q \widetilde{Q} + m' Q' \widetilde{Q'},
\label{superpotentialelectric}
\eea
where the coefficient functions are given by four rotation angles, which
will be discussed in Figure 1, as follows
\bea
\qquad \mu \equiv \tan \theta, 
\qquad \mu' \equiv \tan (\theta'-\theta), \qquad
        \lambda \equiv \sin (\theta-\omega), \qquad \lambda' \equiv
        \sin 
(\theta'-\theta-\omega'). 
\nonu
\eea
Here the adjoint field for $SU(N_c)$ gauge group is denoted by $A$ while
the adjoint field for $SU(N_c')$ gauge group is denoted by $A'$.
The mass terms of these adjoint fields are related to the 
rotation angles of NS5-branes in type IIA brane configuration. 
The couplings of fundamentals with these adjoint
fields are related also to the 
rotation angles of NS5-branes as well as the rotation angles
of D6-branes in type IIA brane configuration.
We add the mass terms for each fundamental flavor.  
The second term in (\ref{superpotentialelectric}) arises 
from the presence of a symmetric flavor $S$ and a conjugate symmetric
flavor $\widetilde{S}$. Except this term and the last
two mass terms, the above superpotential becomes 
the one studied in \cite{BHKL,BH}.
Setting the fields $Q', \widetilde{Q'}, X, \widetilde{X}$ and $A'$ to zero, 
the superpotential becomes the one described in \cite{LLL,Ahn07}.
After integrating out the adjoint fields $A$ and $A'$, 
this superpotential (\ref{superpotentialelectric}) 
at $\theta=\frac{\pi}{2}$ and 
$\theta'=0$ will reduce to the last two mass-deformed terms 
since the coefficient functions 
$\frac{1}{\mu}$
and $\frac{1}{\mu'}$ vanish at this particular rotation angles.
It does not matter whether $\lambda$ or $\lambda'$ vanishes since
eventhough these coefficient functions are not zero, 
$\lambda$- or $\lambda'$-dependent terms all vanish
due to the
$\frac{1}{\mu}$ factor or $\frac{1}{\mu'}$ factor.
For the nonsupersymmetric brane configuration in section 3, we will
consider this particular brane configuration with the constraint 
$\theta=\frac{\pi}{2}$ and 
$\theta'=0$ all the time.

Then what is brane configuration for this gauge theory with given
matter contents? 
It is known that the brane configuration for  
a single gauge group $SU(N_c)$ with a symmetric flavor,
conjugate symmetric flavor and fundamental flavors $S, \widetilde{S},
Q$ and 
$\widetilde{Q}$ is represented by  
the work of \cite{LLL,Ahn07}: 
three NS5-branes, $N_c$ D4-branes, $2N_f$ D6-branes and
orientifold 6-plane where a middle NS5-brane is located.
Now we add the extra two outer NS5-branes, in a ${\bf Z}_2$ symmetric
way due to the O6-plane, to this brane configuration corresponding to 
the first gauge group $SU(N_c)$ and we also put extra $N_c'$  D4-branes 
and extra $N_f'$ $D6$-branes for the second gauge group $SU(N_c')$(and
their mirrors). 

Then
the type IIA brane configuration we are interested in 
consists of five NS5-branes, $N_c$- and $N_c'$- D4-branes 
suspended between them, $2N_f$ and $2N_f'$ D6-branes and orientifold 
6 plane of positive RR charge.
For the negative RR charge, the matter contents of $S$ and
$\widetilde{S}$
are replaced by an antisymmetric and conjugate antisymmetric flavors
$A$ and $\widetilde{A}$, as in \cite{LLL,Ahn07}. 
Let us summarize
the ${\cal N}=1$ supersymmetric electric brane configuration 
we are studying in type IIA string theory as follows:

$\bullet$
First $NS5_{-\theta'}$-brane $(0123vw)$ with $x^6 < 0$

$\bullet$
Second $NS5_{\theta}$-brane $(0123vw)$ with $x^6 < 0$

$\bullet$
Third NS5-brane (012345) with $w=0=x^6$

$\bullet$
Fourth $NS5_{-\theta}$-brane  $(0123vw)$ 
 with $x^6 > 0$

$\bullet$
Fifth $NS5_{\theta'}$-brane  $(0123vw)$ 
with $x^6 > 0$

$\bullet$ First $N_f'$
$D6_{-\omega'}$-branes   $(01237vw)$
with $x^6 < 0$ 

$\bullet$ Second $N_f$
 $D6_{\omega}$-branes  $(01237vw)$ 
with $x^6 < 0$ 

$\bullet$ Third $N_f$
$D6_{-\omega}$-branes   $(01237vw)$ 
 with $x^6 > 0$

$\bullet$ Fourth $N_f'$
$D6_{\omega'}$-branes $(01237vw)$ 
with $x^6 > 0$

$\bullet$ O6-plane  (0123789) with $v=0=x^6$

$\bullet$ $N_c$ D4-branes   (01236) with $v=0=w$

$\bullet$ $N_c'$ D4-branes   (01236) with $v=0=w$

Here we introduce two complex coordinates 
$v \equiv x^4 + i x^5$ and $w \equiv x^8 + i x^9$, as usual,
and the worldvolume $(vw)$ for the rotated branes above corresponds to
the real 2-dimensions spanned in $(v,w)$ plane. The mirrors are located in a 
${\bf Z}_2$ symmetric way.
The $N_c$ D4-branes are suspended between $NS5_{\theta}$-brane and 
$NS5_{-\theta}$-brane while the $N_c'$ D4-branes are suspended between 
$NS5_{-\theta'}$-brane and $NS5_{\theta}$-brane(and their mirrors). 
The convention for the rotated branes is the same as the one used in 
\cite{Ahn07,Ahn07-1,Ahn07-3}.

Let us draw the type IIA brane configuration we describe in Figure 1 and 
we put $N_f$ $D6_{-\omega}$-branes and $N_f'$
$D6_{\omega'}$-branes
in the nonzero $v$ direction for 
nonzero mass terms for the fundamentals(and their mirrors). 
If we are detaching $NS5_{\pm \theta'}$-branes, $D6_{\pm \omega'}$-branes
and $N_c'$ $D4$-branes(and its mirrors), then this brane configuration will 
reduce to the one described in \cite{LLL,Ahn07}.
If we are detaching a middle NS5-brane to the $x^7$ direction, 
then this will lead to the brane configuration considered in \cite{LO,Ahn07-3} 
with the gauge group 
$SO(N_c) \times SU(N_c')$ with fundamentals for each gauge group and 
bifundamentals. 
With O6-plane of negative RR charge instead of having positive RR
charge, 
this process 
will lead to the gauge group 
$Sp(N_c) \times SU(N_c')$ with fundamentals for each gauge group and 
bifundamentals analyzed in \cite{LO,Ahn07-3}. 
If we are detaching all the branes living on the negative $x^6$ region
and O6-plane, then this will become the brane configuration described in 
the work of \cite{BH,BHKL}.

\begin{figure}[ht]
   \epsfxsize=5.0in 
\centerline{\epsffile{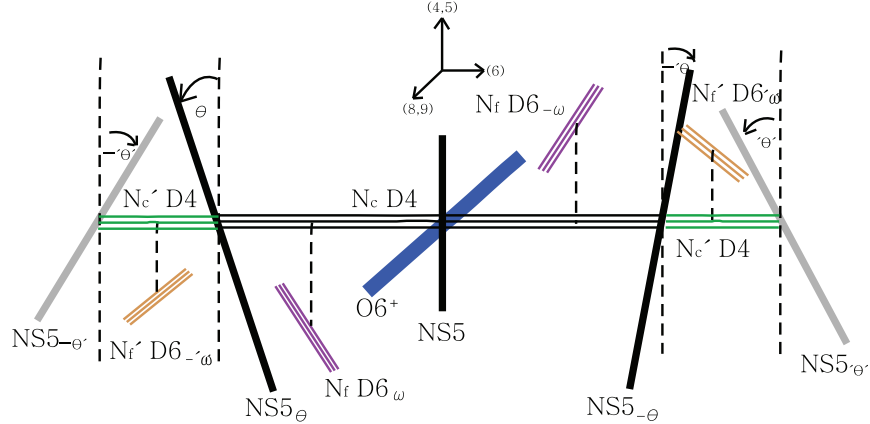}}
   \caption[FIG. \arabic{figure}.]{ 
The ${\cal N}=1$ supersymmetric electric brane configuration with
$SU(N_c) \times SU(N_c')$ gauge group with fundamentals $Q,
\widetilde{Q}, Q'$ and $\widetilde{Q'}$ for each 
gauge group, the bifundamentals $X$ and $\widetilde{X}$ and 
a symmetric flavor $S$ and a
conjugate
symmetric flavor $\widetilde{S}$ for $SU(N_c)$. }
\end{figure}
  
\subsection{Magnetic theory with $SU(\widetilde{N}_c) \times SU(N_c')$
  gauge group}

By brane motion, one gets the Seiberg dual theory
for the gauge group $SU(N_c)$.
From the magnetic brane configuration which is shown in Figure 2 that
is obtained 
by interchanging a set of 
$NS5_{\theta}$-brane and $D6_{\omega}$-branes
and a set of $NS5_{-\theta}$-brane and $D6_{-\omega}$-branes each other,  
the linking number \cite{HW} of $NS5_{\theta=\frac{\pi}{2}}$-brane can
be computed and is
$L_5=\frac{N_f}{2}-\widetilde{N}_c+N_f+N_c'$
when the $N_f$ $D6$-branes are parallel to a middle NS5-brane, as in
the situation of \cite{Ahn07}.
On the other hand, 
the linking number of $NS5_{\theta=\frac{\pi}{2}}$-brane from the
electric brane configuration in Figure 1 can be read off and is 
given by $L_5=-\frac{N_f}{2}+N_c-N_c'$. 
Then the number of dual color $\widetilde{N}_c$, by linking number
conservation, 
is given by
\bea
\widetilde{N}_c = 2(N_f + N_c')-N_c.
\label{dualnumber}
\eea
Compared with the Figure 1, the magnetic brane configuration in Figure
2 has several different features due to the change of the locations
for the branes corresponding to the gauge group $SU(N_c)$. Between the first
$NS5_{-\theta'}$-brane and second $NS5_{-\theta}$-brane, there are
extra $N_f$ $D6_{-\omega}$-branes and newly created $2N_f$ flavor D4-branes
connecting the second $NS5_{-\theta}$-brane and 
$D6_{-\omega}$-branes(and their mirrors) when $\theta \neq
\frac{\pi}{2}$, implying that the matter contents for the gauge group
$SU(N_c')$
will change and this will affect the computation for the coefficient
of beta function below. 
Therefore, these features will provide
the various
interaction terms in the dual magnetic superpotential we will describe later. 

Then the dual magnetic gauge group is  
$SU(\widetilde{N}_c) \times SU(N_c')$ with (\ref{dualnumber}) 
and the matter contents 
are as follows:

$\bullet$ 
$N_f$-chiral multiplets $q$ are  in the
representation $({\bf \widetilde{N}_c}, 1)$,
$N_f$-chiral multiplets $\widetilde{q}$ are in the representation 
$({\bf \overline{\widetilde{N}_c}}, 1)$,
under the gauge group

$\bullet$
$N_f'$-chiral multiplets $Q'$ are  in the
representation $({\bf 1, N_c'
})$, and 
$N_f'$-chiral multiplets $\widetilde{Q'}$ are in  
the representation $({\bf 1,\overline{N_c'}})$,
under the gauge group

$\bullet$
The flavor-singlet field $Y$ is 
in the bifundamental representation $({\bf \widetilde{N}_c, 
\overline{N_c'} })$, 
and its complex conjugate field $\widetilde{Y}$
 is 
in the bifundamental representation $({\bf \overline{\widetilde{N}_c}, 
N_c'})$, 
under the gauge group

$\bullet$ The flavor-singlet field $s$, which is 
in a symmetric tensor representation under the $SU(\widetilde{N}_c)$, 
is in the
representation $({\bf \frac{1}{2} \widetilde{N}_c(\widetilde{N}_c+1),1})$, and
its conjugate field $\widetilde{s}$ is in the 
representation $({\bf \overline{\frac{1}{2} 
\widetilde{N}_c(\widetilde{N}_c+1)},1})$, under the
gauge group

There are also $(N_f+N_c')^2$ gauge-singlets in the first dual gauge group
factor
as follows:

$\bullet$
$N_f$-fields $F'$ are  in the representation $({\bf 1, N_c' })$, 
and its complex conjugate
$N_f$-fields $\widetilde{F'}$ are in the representation 
$({\bf 1, \overline{N_c'} })$, 
under the gauge group

$\bullet$
$N_f^{2}$-fields $M'$ are in the representation $({\bf 1, 1})$ under the
gauge group

$\bullet$
The $N_c^{'2}$-fields 
$\Phi'$ is in the representation $({\bf 1, N_c^{'2}-1}) \oplus ({\bf 1,1
})$ 
under the gauge group  

Moreover, there are additional $N_f(2N_f+1)$ gauge-singlets

$\bullet$
$N_f^{2}$-fields $N'$ are in the representation $({\bf 1, 1})$ under the
gauge group

$\bullet$ $\frac{1}{2} N_f(N_f+1)$-fields $P'$ are 
 in the representation $({\bf 1, 1})$, and its 
conjugate $\frac{1}{2} N_f(N_f+1)$-fields 
 $\widetilde{P'}$  are 
 in the representation $({\bf 1, 1})$,
under the
gauge group

More explicitly, these are represented by $N' \equiv Q \widetilde{S} S
\widetilde{Q},
P' \equiv Q \widetilde{S} Q$ and $\widetilde{P'} \equiv
\widetilde{Q} S \widetilde{Q}$ in terms of fields in electric theory 
explained in \cite{ILS,Ahn07}. 
Although these gauge singlets appear in the dual magnetic
superpotential for the general rotation angles $\theta$ and $\theta'$,
the case $\theta=\frac{\pi}{2}$ we are considering 
does not contain these gauge singlets, as observed in \cite{Ahn07}. 

The coefficient of the beta function of the first dual gauge group factor, 
as done in electric theory, is given by
\bea
b_{SU(\widetilde{N}_c)}^{mag}=3\widetilde{N}_c-N_f-N_c'
-(\widetilde{N}_c+2)
\nonu
\eea
and 
the coefficient of the beta function of the second gauge group factor 
is given by
\bea
b_{SU(N_c')}^{mag}=3N_c'-N_f'-\widetilde{N}_c-N_f-N_c'.
\nonu
\eea
Then both $SU(\widetilde{N}_c)$ and 
$SU(N_c')$ gauge couplings are IR free
by requiring the negativeness of the coefficients of beta function.
One relies on the perturbative calculations at low energy 
for this magnetic IR free region with $b_{SU(\widetilde{N}_c)}^{mag} < 0$ and 
$b_{SU(N_c')}^{mag} < 0$.
It is clear, from the magnetic and electric brane configurations in
Figure 2 and Figure 1,  
that the $SU(N_c')$ fields in the magnetic theory 
are different from those of the electric theory 
\footnote{More explicitly the conditions
  $b_{SU(\widetilde{N}_c)}^{mag} < 0$ and $b_{SU(N_c)} > 0$ imply
that $N_f+N_c' < \frac{2}{3} N_c +\frac{2}{3}$. Also the number of
dual colors $\widetilde{N}_c$ defined as (\ref{dualnumber}) 
should be positive. Then the range for the $N_f$ in the first gauge
group can be written as $ \frac{1}{2} N_c < N_f+N_c' 
< \frac{2}{3} N_c +\frac{2}{3}$. 
Since $b_{SU(N_c')}-b_{SU(N_c')}^{mag} = 3(N_c'+N_f)-2N_c < 0$,
if we require that $b_{SU(N_c')}^{mag} < 0$ which is equivalent to 
$N_c-3N_f < N_f'$, then the electric description of $SU(N_c')$ is IR
free because $b_{SU(N_c')} < 0$.
At high energy, $SU(N_c')$ theory is strongly coupled while $SU(N_c)$
theory
is UV free. At the scale $\Lambda_1$, the $SU(N_c)$ theory is strongly
coupled and the Seiberg duality occurs. All the running couplings are
changed by this duality and all the coefficients of beta functions, 
$b_{SU(\widetilde{N}_c)}^{mag}$ and $b_{SU(N_c')}^{mag}$
become negative. Then at energy scale lower than $\Lambda_1$, 
the theory is weakly coupled. When 
$b_{SU(N_c')}^{mag} < b_{SU(\widetilde{N}_c)}^{mag} < 0$, 
the one loop computation is reliable with $\Lambda_1 << \Lambda_2$. 
When 
$b_{SU(\widetilde{N}_c)}^{mag} < b_{SU(N_c')}^{mag} < 0$, 
the requirement that 
$SU(N_c')^{mag}$ theory is less coupled than the 
$SU(\widetilde{N}_c)^{mag}$ at the supersymmetry breaking scale $\mu$
provides a stronger constraint on $\Lambda_2$. It is not enough 
to choose it higher than $\Lambda_1$ simply.
Then under the constraint, $\Lambda_2 >> \left(
  \frac{\Lambda_1}{\mu}\right)^b
\Lambda_1$
where $b \equiv 
\frac{b_{SU(\widetilde{N}_c)}^{mag}-b_{SU(N_c')}^{mag}}{b_{SU(N_c')}}$, 
one can ignore the contribution from the gauge coupling of 
$SU(N_c')^{mag}$ at the supersymmetry breaking scale and one relies on
the one loop computation. 
See
the ref. \cite{AGM07} for the relevant discussions in the context of 
quiver gauge theory.  In particular, the appendix B of \cite{AGM07}.}.

The dual magnetic superpotential 
\footnote{
Although gauging the $SU(\widetilde{N}_c)$ does not affect the supersymmetry
breaking vacua which will be discussed in next section, it leads to
the supersymmetry vacua. We integrate out the bifundamentals $Y$ and $\widetilde{Y}$
in such a way that the gauge group $SU(\widetilde{N}_c)$ is not broken 
by the fields $Y$ and $\widetilde{Y}$, as in meta-stable state, so 
$< Y > = 0 = < \widetilde{Y} >$ in (\ref{vacuum1}). 
For nonzero vacuum expectation values for $M'$, this
superpotential gives the $SU(\widetilde{N}_c)$  ``flavors''
$q \widetilde{s}$ and $s \widetilde{q}$, the mass $<M'>$. Below the energy scale 
$<M'>$, one can integrate out these massive flavors using the equations
of motion $< q \widetilde{s}> = 0 = < s \widetilde{q} >$. Then the low energy theory
is given by $SU(\widetilde{N}_c)$ pure Yang-Mills theory and the
corresponding scale matching condition connecting between the low
energy scale $\Lambda_L$ and the macroscopic scale
$\widetilde{\Lambda}$ 
can be computed. Then the low energy theory has a superpotential term 
 which is proportional to $\left( 
\widetilde{\Lambda}^{2\widetilde{N}_c-N_f-N_c'-2} \mbox{det} M' 
\right)^{\frac{1}{\widetilde{N}_c}}$ plus $m M'$. Using this
dynamically generated superpotential the vacuum expectation value for
$M'$
is obtained in the supersymmetric vacuum.
There is no
conserved $U(1)_R$ symmetry because it is anomalous under the gauged 
$SU(\widetilde{N}_c)$ in the sense that the determinant term above
breaks it explicitly. 
Therefore, the $U(1)_R$ symmetry returns 
an ``approximate'' accidental symmetry of the IR theory. See the
ref. \cite{NS} for the discussion on the relation between the R-symmetry
breaking and supersymmetry breaking and also the refs. \cite{IS,ISS07} on the recent
revival on this subject. The small parameter of \cite{ISS07}
corresponds to the above
$\widetilde{\Lambda}^{(2\widetilde{N}_c-N_f-N_c'-2)/\widetilde{N}_c}$ 
with negative exponent.   }
for massless fundamental flavors
 $Q'$
and $\widetilde{Q'}$(i.e., $m'=0$) and massive fundamental flavors 
$Q$ and $\widetilde{Q}$ is given by
\bea
W_{dual}= \left(M' q \widetilde{s} s \widetilde{q} +  m M' \right)+ 
\widetilde{Y} \widetilde{F'} q +
Y \widetilde{q} F' + \Phi' Y \widetilde{Y} +\left( \Phi^{'2} + \cdots \right)
\label{dualsup}
\eea
where the mesons are given in terms of fields in the 
electric theory(See also the relevant works found in \cite{Ahn07-2,Ahn07-3}) 
\bea
M' \equiv Q \widetilde{Q}, \qquad 
F' \equiv \widetilde{X} Q, \qquad 
\widetilde{F'} \equiv X \widetilde{Q}, \qquad
\Phi' \equiv X \widetilde{X}.
\nonu
\eea
Here the last piece $\left( \Phi^{'2} + \cdots \right)$ in $W_{dual}$
above is coming from the superpotential (\ref{superpotentialelectric}) 
in an electric theory and  contains also $ N' q \widetilde{q} + P' q 
\widetilde{s} q +\widetilde{P'} \widetilde{q} s \widetilde{q}$ for
 the general rotation angles $\theta, \theta', \omega$ and $\omega'$.  
When both $\theta=\frac{\pi}{2}$ and $\theta'=0$, this piece will vanish and the
superpotential consists of the first five terms in (\ref{dualsup})
that are relevant part for the meta-stable brane configuration next section.
The strings stretching between the $N_f$ $D6_{\omega}$-branes and 
$N_c'$ D4-branes lead to the gauge theory objects for 
the additional $N_f$ $SU(N_c')$ fundamentals 
$F'$ and the additional $N_f$ $SU(N_c')$ antifundamentals $\widetilde{F'}$.
The fluctuations of the singlet $\Phi'$ correspond to the motion of 
$N_c'$ D4-branes suspended two NS5-branes(and its mirrors).
The fluctuations of the singlet $M'$ correspond to the motion of 
additional $N_f$-flavor 
D4-branes suspended between D6-branes and NS5-brane(and its mirrors).

\begin{figure}[ht]
   \epsfxsize=5.0in 
\centerline{\epsffile{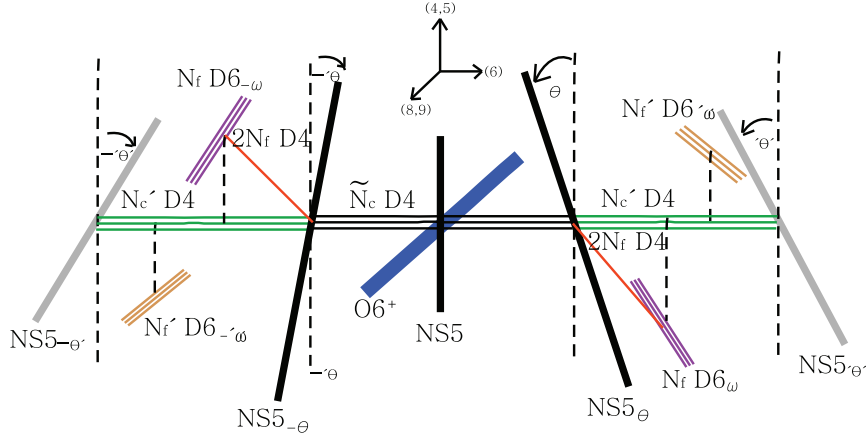}}
   \caption[FIG. \arabic{figure}.]{ 
The ${\cal N}=1$ supersymmetric magnetic brane configuration with
$SU(\widetilde{N}_c=2N_f+2N_c'-N_c) 
\times SU(N_c')$ gauge group with fundamentals $q,
\widetilde{q}, Q'$ and $\widetilde{Q'}$ for each 
gauge group, the bifundamentals $Y$ and $\widetilde{Y}$ and 
a symmetric flavor $s$ and a
conjugate
symmetric flavor $\widetilde{s}$ for $SU(\widetilde{N}_c)$ and various
gauge singlets. 
When $\theta=\frac{\pi}{2}$, the number of newly created flavor D4-branes
connecting $D6_{\omega}$-branes and $NS5_{\theta=\frac{\pi}{2}}$ is
reduced to $N_f$, not $2N_f$ as observed in \cite{Ahn07}.
Compared with the electric brane configuration in 
Figure 1, the second NS5-brane with $N_f$ D6-branes and the fourth NS5-brane 
with other $N_f$ D6-branes are interchanged along $x^6$-direction each other.}
\end{figure}

\subsection{Magnetic theory with $SU(N_c) \times SU(\widetilde{N}_c')$
  gauge group}

In this subsection, we consider the other magnetic theory. 
By brane motion, one gets the Seiberg dual theory
for the second gauge group $SU(N_c')$.
From the magnetic brane configuration which is shown in Figure 3 obtained 
by interchanging a set of 
$NS5_{\theta'}$-brane and $D6_{\omega'}$-branes
and $NS5_{-\theta}$-brane,
the linking number of $NS5_{-\theta=\frac{\pi}{2}}$-brane is
given by $L_5=\frac{N_f'}{2}-\widetilde{N}_c'$ as long as 
$D6_{\omega'}$-branes are not parallel to $NS5_{-\theta}$-brane.
Moreover, the linking number of $NS5_{-\theta=\frac{\pi}{2}}$-brane 
from the electric brane configuration in Figure 1 is 
given by $L_5=-\frac{N_f'}{2}+N_c'-N_c$.
Then the number of dual color $\widetilde{N}_c$, by linking number
conservation, 
is given by
\bea
\widetilde{N}_c' = N_f' + N_c-N_c'.
\label{num}
\eea
Compared with the Figure 1, the magnetic brane configuration in Figure
3 has several different features. Between the second
$NS5_{-\theta'}$-brane and third NS5-brane, there are
extra $N_f'$ $D6_{-\omega'}$-branes and newly created $N_f'$ D4-branes
connecting the second $NS5_{-\theta'}$-brane and 
$D6_{-\omega'}$-branes(and their mirrors), implying that the $SU(N_c)$
fields will change and affect for the computation of coefficient of
beta function below. 
So this will provide
the interaction terms, that did not appear in the electric theory,
in the dual magnetic superpotential. 

Then the dual magnetic gauge group is  
$SU(N_c) \times SU(\widetilde{N}_c')$ with (\ref{num}) and 
the matter contents 
are as follows:

$\bullet$
$N_f$-chiral multiplets $Q$ are  in the
representation $({\bf N_c,1
})$, and 
$N_f$-chiral multiplets $\widetilde{Q}$ are in  
the representation $({\bf \overline{N_c}, 1})$,
under the gauge group

$\bullet$ 
$N_f'$-chiral multiplets $q'$ are  in the
representation $({\bf 1,\widetilde{N}_c'})$,
$N_f'$-chiral multiplets $\widetilde{q'}$ are in the representation 
$({\bf 1, \overline{\widetilde{N}_c'}})$,
under the gauge group

$\bullet$
The flavor-singlet field $Y$ is 
in the bifundamental representation $({\bf N_c, 
\overline{\widetilde{N}_c'} })$, 
and its complex conjugate field $\widetilde{Y}$
 is 
in the bifundamental representation $({\bf \overline{N_c}, 
\widetilde{N}_c'})$, 
under the gauge group

$\bullet$ The flavor-singlet field $S$, which is 
in a symmetric tensor representation under the $SU(N_c)$, is in the
representation $({\bf \frac{1}{2} N_c(N_c+1),1})$, and
its conjugate field $\widetilde{S}$ is in the 
representation $({\bf \overline{\frac{1}{2} N_c(N_c+1)},1})$, under the
gauge group

There are $(N_f'+N_c)^2$ gauge-singlets in the second dual gauge group
factor
as follows:

$\bullet$
$N_f'$-fields $F$ are  in the representation $({\bf N_c,1 })$, 
and its complex conjugate
$N_f'$-fields $\widetilde{F}$ are in the representation 
$({\bf \overline{N_c},1 })$, 
under the gauge group

$\bullet$
$N_f^{'2}$-fields $M$ are in the representation $({\bf 1, 1})$ under the
gauge group

$\bullet$
The $N_c^2$-fields 
$\Phi$ is in the representation $({\bf N_c^2-1, 1}) \oplus ({\bf 1,1
})$ 
under the gauge group  

The coefficient of the beta function of the first gauge group factor 
is given by
\bea
b_{SU(N_c)}^{mag}=3N_c-N_f-\widetilde{N}_c'-N_f'-N_c -
(N_c+2)
\nonu
\eea
and 
the coefficient of the beta function of the second gauge group factor 
is given by
\bea
b_{SU(\widetilde{N}_c')}^{mag}=3\widetilde{N}_c'-N_f'-N_c.
\nonu
\eea
It is evident that the $SU(N_c)$ fields in the magnetic theory in
Figure 3 
are different from those of the electric theory in Figure 1
 \footnote{Now the conditions
  $b_{SU(\widetilde{N}_c')}^{mag} < 0$ and $b_{SU(N_c')} > 0$ imply
that $N_f'+N_c < \frac{3}{2} N_c' $. Also the number of
dual colors $\widetilde{N}_c'$ defined as (\ref{num}) 
should be positive. Then the range for the $N_f'$ in the second gauge
group can be written as $ N_c' < N_f'+N_c 
< \frac{3}{2} N_c'$. 
 The condition
$b_{SU(N_c)}^{mag} < 0$ implies 
 $ N_c'-2N_f'-2 < N_f$.
The $b_{SU(N_c)}$ can be IR free or UV free in the electric description. 
If the former where $b_{SU(N_c)} < 0$ happens, then one can analyze 
the method in the footnote 1 exactly.
When 
$b_{SU(N_c)}^{mag} < b_{SU(\widetilde{N}_c')}^{mag} < 0$, 
the one loop computation is reliable with $\Lambda_2 << \Lambda_1$. 
When 
$b_{SU(\widetilde{N}_c')}^{mag} < b_{SU(N_c)}^{mag} < 0$, 
under the constraint, $\Lambda_1 >> \left(
  \frac{\Lambda_2}{\mu}\right)^b
\Lambda_2$
where $b \equiv 
\frac{b_{SU(\widetilde{N}_c')}^{mag}-b_{SU(N_c)}^{mag}}{b_{SU(N_c)}}$, 
one can ignore the contribution from the gauge coupling of 
$SU(N_c)^{mag}$ at the supersymmetry breaking scale and one relies on
the one loop computation. 
If the latter happens where 
$b_{SU(N_c)} > 0$, then for the case where the 
$SU(N_c)^{mag}$ theory becomes more IR free than the 
$SU(\widetilde{N}_c')^{mag}$, in other words, 
$b_{SU(N_c)}^{mag} < b_{SU(\widetilde{N}_c')}^{mag} < 0$  
after Seiberg duality the coupling of the $SU(N_c)^{mag}$ becomes more
smaller than the coupling of $SU(\widetilde{N}_c')^{mag}$ along the
flow to the low energy. Then the one loop computation is reliable with 
$\Lambda_1 << \Lambda_2$. 
For the case  where the 
$SU(N_c)^{mag}$ theory becomes less IR free than the 
$SU(\widetilde{N}_c')^{mag}$, in other words, 
$ b_{SU(\widetilde{N}_c')}^{mag} < b_{SU(N_c)}^{mag} < 0$, 
under the strong constraint, $\Lambda_1 << \left(
  \frac{\Lambda_2}{\mu}\right)^b
\Lambda_2 << \Lambda_2$
where $b$ is the same as above, 
one can ignore the contribution from the gauge coupling of 
$SU(N_c)^{mag}$ at the supersymmetry breaking scale and one relies on
the one loop computation. 
}.
Since $b_{SU(N_c)}-b_{SU(N_c)}^{mag} > 0$, the $SU(N_c)$ theory is more
asymptotically free than the $SU(N_c)^{mag}$ theory \cite{BIWW}.

The dual magnetic superpotential 
\footnote{
Although gauging the $SU(\widetilde{N}_c')$ does not affect the supersymmetry
breaking vacua which will be discussed in next section, it leads to
the supersymmetry vacua. For nonzero vacuum expectation values for $M$, this
superpotential gives the $SU(\widetilde{N}_c')$ fundamental flavors
$q'$ and $\widetilde{q'}$, the mass $<M>$. Below the energy scale 
$<M>$, one can integrate out these massive flavors using the equations
of motion $< q' > = 0 = < \widetilde{q'} >$. Then the low energy theory
is given by $SU(\widetilde{N}_c')$ pure Yang-Mills theory and the
corresponding scale matching condition connecting between the low
energy scale $\Lambda_L$ and the macroscopic scale
$\widetilde{\Lambda}$ 
can be computed. Then the low energy theory has a superpotential term 
 which is proportional to $\left( 
\widetilde{\Lambda}^{3\widetilde{N}_c'-N_f'-N_c} \mbox{det} M 
\right)^{\frac{1}{\widetilde{N}_c'}}$ plus $m' M$. There is no
conserved $U(1)_R$ symmetry because it is anomalous under the gauged 
$SU(\widetilde{N}_c')$ in the sense that the determinant term above
breaks it explicitly. Remember that the $R$ charges for the fields are
as follows: $R(Y)=R(q')=\frac{N_c'}{N_f'+N_c}$ and $R(M)=R(F)=R(\Phi)=
2-\frac{2N_c'}{N_f'+N_c}$. Therefore, the $U(1)_R$ symmetry returns an
 ``approximate'' accidental symmetry of the IR theory.  } 
for massless fundamental flavors
$Q$
and $\widetilde{Q}$(i.e., $m=0$) and massive fundamental flavors $Q'$
and $\widetilde{Q'}$ is 
\bea
W_{dual}'= \left( M q' \widetilde{q'}  + m' M \right) + 
Y \widetilde{F} q' +
\widetilde{Y} \widetilde{q'} F + \Phi Y \widetilde{Y} 
+ \left( \Phi^2 + \cdots \right)
\label{dualdual}
\eea
where  the mesons are given by
\bea
M \equiv Q' \widetilde{Q'}, \qquad 
F \equiv X Q', \qquad 
\widetilde{F} \equiv \widetilde{X} \widetilde{Q'}, \qquad
\Phi \equiv X \widetilde{X}.
\nonu
\eea
Here the last piece $\left( \Phi^2 + \cdots \right)$ in $W_{dual}$
is coming from the superpotential (\ref{superpotentialelectric}) 
in an electric theory for
 the general rotation angles $\theta, \theta', \omega$ and $\omega'$.  
When $\theta=\frac{\pi}{2}$ and $\theta'=0$, this will vanish and the
superpotential consists of the first five terms in (\ref{dualdual}).
As we observed, the
presence of  
$N_f'$ $D6_{\omega'}$-branes and $N_f'$ D4-branes give rise to 
the gauge-singlets. That is,
the strings stretching between the $N_f'$ $D6_{\omega'}$-branes and 
$N_c$ D4-branes lead to the additional $N_f'$ $SU(N_c)$ fundamentals 
$F$ and additional $N_f'$ $SU(N_c)$ antifundamentals $\widetilde{F}$.
The fluctuations of the singlet $\Phi$ correspond to the motion of 
$N_c$ D4-branes suspended two NS5-branes(and its mirrors).
The fluctuations of the singlet $M$ correspond to the motion of 
additional $N_f'$ flavor 
D4-branes suspended between D6-branes and NS5-brane(and its mirrors).

\begin{figure}[ht]
   \epsfxsize=5.0in 
\centerline{\epsffile{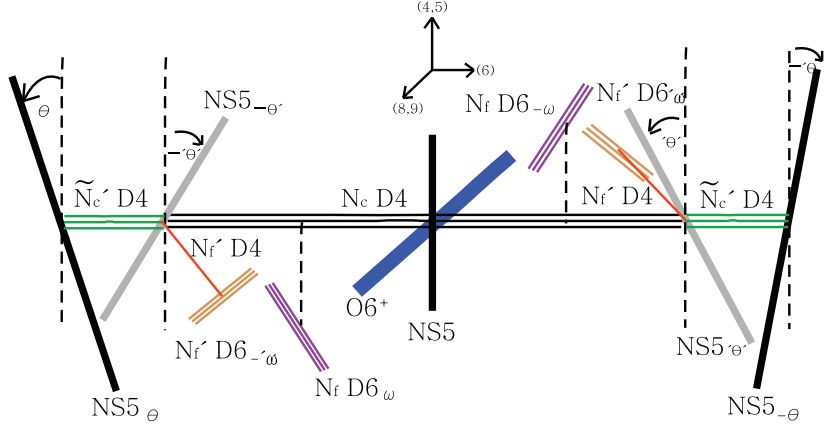}}
   \caption[FIG. \arabic{figure}.]{ 
The ${\cal N}=1$ supersymmetric magnetic brane configuration with
$SU(N_c) \times SU(\widetilde{N}_c'=N_f'+N_c-N_c')$ 
gauge group with fundamentals $Q,
\widetilde{Q}, q'$ and $\widetilde{q'}$ for each 
gauge group, the bifundamentals $Y$ and $\widetilde{Y}$ 
and 
a symmetric flavor $S$ and a
conjugate
symmetric flavor $\widetilde{S}$ for $SU(N_c)$
and various
gauge singlets. 
Compared with Figure 1, the first NS5-brane and the 
second NS5-brane with $N_f'$ D6-branes 
are interchanged along $x^6$ direction each other(and
its mirrors).  }
\end{figure}

\section{Nonsupersymmetric meta-stable brane configuration }

Based on the magnetic brane configurations we have found in previous
section,
we describe the nonsupersymmetric brane configurations by
recombination of flavor D4-branes and color D4-branes and splitting
between those flavor D4-branes and the remnant of flavor D4-branes
which does not participate in the  recombination process.

\subsection{ When 
the magnetic gauge group is $SU(\widetilde{N}_c) \times SU(N_c')$}

In this case, the dual magnetic superpotential is given by the first
five terms of 
(\ref{dualsup}) at $\theta=\frac{\pi}{2}$ and $\theta'=0$.
The dual quarks $q$ and $\widetilde{q}$ are fundamental 
$({\bf \widetilde{N}_c,1})$ and
antifundamental $({\bf \overline{\widetilde{N}_c}},1)$ 
for the gauge group indices and antifundamentals for
the flavor indices. 
The flavor-singlet fields $s$ and $\widetilde{s}$ are symmetric  
$({\bf \frac{1}{2} \widetilde{N}_c(\widetilde{N}_c+1),1})$ and conjugate
symmetric tensor 
 $({\bf \overline{\frac{1}{2} 
\widetilde{N}_c(\widetilde{N}_c+1)},1})$
for the gauge group indices respectively. The quantity $q
\widetilde{s} s \widetilde{q}$ has a rank $\widetilde{N}_c$ and the
mass matrix $m$ has a rank $N_f$. Then the F-term equation for $M'$
cannot be satisfied if the rank $N_f$ exceeds $\widetilde{N}_c$ and
the supersymmetry is broken.

The classical moduli space of vacua can be obtained from F-term
equations and one gets
\bea
 q \widetilde{s} s \widetilde{q} +  m & = & 0, \qquad
\widetilde{s} s \widetilde{q} M' + \widetilde{Y} \widetilde{F'}   =  0, \nonu \\
s \widetilde{q} M' q & = & 0, \qquad
\widetilde{q} M' q \widetilde{s}  =  0, \nonu \\
  M' q \widetilde{s} s + F' Y & = & 0, \qquad
\widetilde{F'} q + \Phi' Y  =  0,
\nonu \\
q \widetilde{Y} & = & 0, \qquad
\widetilde{q} F' + \widetilde{Y} \Phi'  =  0, \nonu \\
Y \widetilde{q} & = & 0, \qquad
Y \widetilde{Y}  =  0. 
\nonu
\eea
Some of F-term equations are satisfied if one takes the zero vacuum
expectation values for the fields $Y, \widetilde{Y}, F'$ and 
$\widetilde{F'}$. 
Then, it is easy to see that 
\bea
s \widetilde{q} M' =0= M' q \widetilde{s}, \qquad
 q \widetilde{s} s \widetilde{q} +  m  =  0.
\nonu 
\eea
Then the solutions can be written as
\bea
<q \widetilde{s}>  & = &  \left(
\begin{array}{c}
\sqrt{m} e^{\phi} {\bf 1}_{\widetilde{N}_c}  \\
0
\end{array}
\right),  
<s \widetilde{q}> =
 \left(
\begin{array}{cc}
\sqrt{m} e^{-\phi}  {\bf 1}_{\widetilde{N}_c}   &
0
\end{array}
\right), 
<M'>  =
 \left(
\begin{array}{cc}
0  & 0 
 \\
0 & \Phi_0  {\bf 1}_{N_f-\widetilde{N}_c} 
\end{array}
\right)
\nonu \\
<Y> & = & <\widetilde{Y}> = <F'> = <\widetilde{F'}>= 0.
\label{vacuum1}
\eea
Let us expand around on a point on (\ref{vacuum1}), as done in
\cite{ISS}. 
Then the remaining relevant terms of superpotential are given by
\bea
W_{dual}^{rel} & = &  \Phi_0 \left( \delta \hat{\varphi}  
\; \delta \hat{\widetilde{\varphi}} + m \right) +
  \delta Z \; \delta \hat{\varphi} \; s_0 \; \widetilde{q}_0 
+ \delta \widetilde{Z} \; q_0 \; \widetilde{s}_0 \;
\delta \hat{\widetilde{\varphi}}
\nonu
\eea
by following the fluctuations for the various fields in \cite{Ahn07}.
Note that there exist three kinds of terms, 
the vacuum  $<q>$ multiplied by 
$\delta \widetilde{Y} \delta \widetilde{F'}$,  
the vacuum  $<\widetilde{q}>$ multiplied by $\delta F' 
\delta Y$, and 
the vacuum  $<\Phi'>$ multiplied by $\delta Y 
\delta \widetilde{Y}$.
By redefining these as 
 $\delta \hat{\widetilde{Y}} \delta \hat{\widetilde{F'}}, \delta \hat{F'} 
\delta \hat{Y}$, and $\delta \hat{Y} 
\delta \hat{\widetilde{Y}}$
respectively, they do not enter the 
contributions for the one loop result, up to quadratic order. 
As done in \cite{Ahn07}, the defining function ${\cal F}(v^2)$ can be
computed
and using the equation (2.14) of \cite{Shih} 
of $m_{\Phi_0}^2$ and ${\cal F}(v^2)$, one gets 
that $m_{\Phi_0}^2$ will contain $(\log 4 -1) > 0$ implying that these
are stable.

Let us recombine $\widetilde{N}_c$  flavor D4-branes among the
additional
$N_f$ flavor 
D4-branes with those connecting NS5'-brane(coming from
$NS5_{-\theta}$-brane) 
and NS5-brane and then push
them in the $+v$ direction from the magnetic brane configuration 
in Figure 2, as done in \cite{OO,FGU,BGHSS}. 
Due to the presence of middle NS5-brane, 
this procedure, pushing into the $+v$ diretion, is possible.
This is different feature, compared with the one in \cite{Ahn07-3}
where there was no meta-stable brane configuration, 
when we take the Seiberg dual for the gauge group $SO(N_c)$,
since there was no extra NS5-brane, unlike to the present case.   
Of course, the mirrors will move $-v$
direction due to the presence of O6-plane. There are no color
D4-branes connecting NS5'-brane and NS5-brane and there exist only 
$(N_f-\widetilde{N}_c)$ flavor D4-branes connecting D6-branes and
NS5'-brane(and their mirrors) that are misaligned to the above
$\widetilde{N}_c$
flavor D4-branes.

Then, the minimal energy supersymmetry breaking brane configuration 
is given by Figure 4 where one sees 
a misalignment between the additional flavor
D4-branes. If we are detaching $N_c'$ D4-branes, $N_f'$
D6-branes and NS5-brane(coming from $NS5_{\theta'}$-brane)(and its
mirrors), 
then this brane configuration leads to the one described in \cite{Ahn07}.
If we are detaching all the branes living on the positive $x^6$ region
and O6-plane, then this will look like  
the brane configuration of \cite{Ahn07-3} with the product gauge group
of unitary group shown in Figure 3 of \cite{Ahn07-3}. The difference 
between these two appears in the the left NS5-brane: in Figure 4, it
is NS5-brane while in Figure 3 of \cite{Ahn07-3}, it is given by 
NS5'-brane.

\begin{figure}[ht]
   \epsfxsize=5.0in 
\centerline{\epsffile{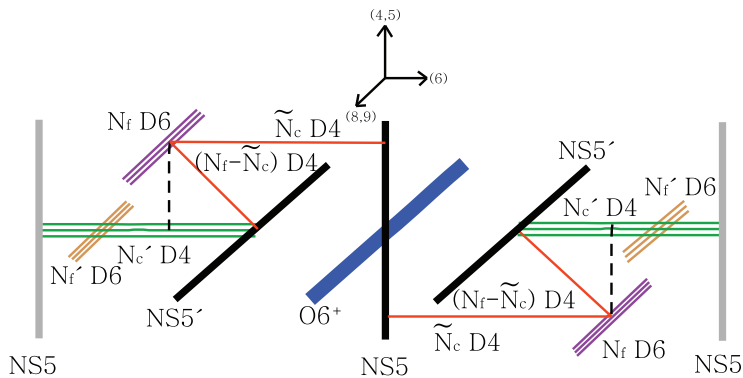}}
   \caption[FIG. \arabic{figure}.]{ 
The nonsupersymmetric minimal energy brane configuration of 
$SU(\widetilde{N}_c=2N_f +2N_c'-N_c) \times SU(N_c')$ 
 gauge group with fundamentals $q,
\widetilde{q}, Q'$ and $\widetilde{Q'}$ for each 
gauge group, the bifundamentals $Y$ and $\widetilde{Y}$ and 
a symmetric flavor $s$ and a
conjugate
symmetric flavor $\widetilde{s}$ for $SU(\widetilde{N}_c)$ and various
gauge singlets. 
We consider the massless case of $Q'$ and $\widetilde{Q'}$.
 Compared with Figure 2, there is a misalignment of the additional $N_f$ flavor
 D4-branes. Some of them are connecting to NS5'-brane and the other to
NS5-brane. In this figure, the rotation angle $\theta$ of Figure 2 
is $\frac{\pi}{2}$ while the rotation angle $\theta'$ is 0.}
\end{figure}

In \cite{Witten,LL}, the Riemann surface 
describing a set of NS5-branes 
with D4-branes suspended between them and 
in a background space of $x t = (-1)^{N_f+N_f'} v^4
\prod_{k=1}^{N_f} (v^2 -e_k^2) \prod_{l=1}^{N_f'} (v^2 - e_l^2)$
was found.
Since we are dealing with five NS5-branes, the magnetic M5-brane 
configuration in Figure 2 with equal mass for $Q$ and $\widetilde{Q}$ and massless 
for $Q'$ and $\widetilde{Q'}$ 
can be characterized by the following quintic equation for $t$ 
as follows:
\bea
& & t^5 + \left[v^{N_c'} \right] t^4 + \left[ v^{\widetilde{N_c}+N_f'}
 (v -m)^{N_f} \right] t^3 + \left[ (-1)^{\widetilde{N}_c}
v^{\widetilde{N}_c +2N_f'+ 2} 
(v -m)^{2N_f} \right] t^2 \nonu \\
& & + \left[ (-1)^{N_c'} v^{N_c' +3N_f'+6}  
(v -m)^{3N_f} \right] t + \left[(-1)^{N_f+N_f'}
v^{10+5N_f'}  (v -m)^{4N_f} 
 (v +m)^{N_f}\right] = 0.
\nonu
\eea
The polynomial $g_1(v)$ \cite{LL} appearing in the coefficient $t^4$ 
in $v$ of degree $N_c'$ given by the number of
D4-branes suspended between the first and second NS5-branes from
Figure 4 
has a highest power $N_c'$ of $v$. The polynomial $g_4(v)$ appearing in the
coefficient
$t$ in $v$ of degree 
$N_c'$ given by the number of D4-branes suspended between the fourth
and fifth NS5-branes from Figure 4 can be expressed in terms of 
$g_1(v)$ with a replacement $v \rightarrow -v$.
Similarly, 
the polynomial $g_2(v)$ appearing in the coefficient $t^3$ 
in $v$ of degree $\widetilde{N}_c$ given by the number of 
D4-branes suspended between the second and third NS5-branes from
Figure 2 
has a highest power $\widetilde{N}_c$ of $v$ and the polynomial 
$g_3(v)$ appearing in the coefficient $t^2$ in $v$ of degree
$\widetilde{N}_c$
given by the number of D4-branes suspended between  
the third and fourth NS5-branes from Figure 2 can be written as 
$g_2(v)$ with a replacement $v \rightarrow -v$.
We also used the symmetry in 
other polynomials $J_i(v)$ where $i=1,2,3,4$ 
representing 
the contribution to the above space relating to the complex variables 
$x, t$ and $v$ from D6-brane charge sources, i.e., D6-branes and O6-plane
between $i$-th an $(i+1)$-th NS5-branes.
Note that $J_2(v) = J_3(v) =v^2$ and $\prod_{i=1}^{4} J_i(v)=
 (-1)^{N_f+N_f'} v^{2N_f'+4} (v^2 -m^2)^{N_f}$.
Moreover, there exist relations $J_3(v)=J_2(-v)$ and $J_4(v)=J_1(-v)$ 
due to the ${\bf Z}_2$ symmetry of O6-plane.

At nonzero string coupling constant, 
the NS5-branes bend due to their interactions with the D4-branes and
D6-branes.
Now the asymptotic regions of various NS5-branes 
can be determined by reading off the first two terms of the quintic
curve above giving the
$NS5_L$-brane
asymptotic region, next two terms giving 
$NS5'_L$-brane asymptotic region, next two terms
giving $NS5_M$-brane asymptotic region, next two terms 
giving $NS5'_R$-brane asymptotic region, and 
final two terms giving $NS5_R$-brane asymptotic region.
Then the behavior of the supersymmetric M5-brane curves can be
summarized 
as follows:

1. $v \rightarrow \infty$ limit implies
\bea
w & \rightarrow & 0, \quad y \sim    v^{N_c'} + \cdots \quad
\mbox{$NS5_{L}$ 
asymptotic region}, \nonu \\
w & \rightarrow  & 0, \quad y \sim    
v^{N_f+N_f'+2} + \cdots \quad
\mbox{$NS5_{M}$ asymptotic region}, 
\nonu \\
w & \rightarrow  & 0, \quad y \sim    
v^{2N_f+2N_f'-N_c'+4} + \cdots \quad
\mbox{$NS5_{R}$ asymptotic region}.   
\nonu
\eea

2.  $w \rightarrow \infty$ limit implies
\bea
v & \rightarrow &   +m, \quad 
y \sim  w^{\widetilde{N}_c-N_c'+N_f+N_f'}
 +\cdots
\quad \mbox{$NS5_{L}'$ asymptotic region}, 
\nonu
\\
v & \rightarrow &  -m, \quad  
y \sim w^{N_c'-\widetilde{N}_c+N_f+N_f'+4}
+\cdots
\quad \mbox{$NS5_{R}'$ asymptotic region}. 
\nonu
\eea

The two $NS5_{L,R}'$-branes 
are moving in the $\pm v$ direction respectively holding everything 
else fixed instead of moving D6-branes in the $\pm v$ direction.
The corresponding mirrors of D4-branes are moved appropriately.
The harmonic function in the Tau-NUT space, sourced by $2N_f$
D6-branes, O6-plane and $2N_f'$ D6-branes, can be determined once we
fix the $x^6$ position for these branes. Then the first order
differential equation for the $g(s)$ \cite{BGHSS} where the absolute value of
$g(s)$ is equal to the absolute value of $w$ can be solved exactly with the
appropriate boundary conditions on $NS5_L'$ or $NS5_R'$ asymptotic
region from above classification 2. Since the extra terms in the harmonic
function contribute to the $g(s)$ as a multiplication factor,   
the contradiction with the correct statement that 
$y$ should vanish only if $v=0$,
implies that there exists the instability from a new M5-brane mode at
some point from the transition of SQCD-like theory description 
to M-theory description.

\subsection{ When 
the magnetic gauge group is $SU(N_c) \times SU(\widetilde{N}_c')$}

In this case, the dual magnetic superpotential is given by the first
five terms of 
(\ref{dualdual}) at $\theta=\frac{\pi}{2}$ and $\theta'=0$.
The dual quarks $q'$ and $\widetilde{q'}$ are 
fundamental $({\bf 1, \widetilde{N}_c'})$ and
antifundamental $({\bf 1, \overline{\widetilde{N}_c'}})$ 
for the gauge group indices and antifundamentals for
the flavor indices. 
The quantity $q'
 \widetilde{q'}$ has a rank $\widetilde{N}_c'$ and the
mass matrix $m'$ has a rank $N_f'$. Then the F-term equation for $M$
cannot be satisfied if the rank $N_f'$ exceeds $\widetilde{N}_c'$ and
the supersymmetry is broken.

The classical moduli space of vacua can be obtained from F-term
equations and one gets
\bea
 q'  \widetilde{q'} +  m' & = & 0, \qquad
\widetilde{q'} M + Y \widetilde{F}   =  0, \nonu \\
  M q'  + F \widetilde{Y} & = & 0, \qquad
\widetilde{F} q' +  \widetilde{Y} \Phi  =  0,
\nonu \\
q' Y & = & 0, \qquad
\widetilde{q'} F +  \Phi Y  =  0, \nonu \\
\widetilde{Y} \widetilde{q'} & = & 0, \qquad
Y \widetilde{Y}  =  0. 
\nonu
\eea
Other F-term equations are satisfied if one takes the zero vacuum
expectation values for the fields $Y, \widetilde{Y}, F$ and 
$\widetilde{F}$.
Then, it is easy to see that 
\bea
\widetilde{q'} M =0= M q', \qquad
 q'  \widetilde{q'} +  m'  =  0.
\nonu 
\eea
Then the solutions can be written as
\bea
<q' >  & = &  \left(
\begin{array}{c}
\sqrt{m} e^{\phi} {\bf 1}_{\widetilde{N}_c'}  \\
0
\end{array}
\right),  
< \widetilde{q'}> =
 \left(
\begin{array}{cc}
\sqrt{m} e^{-\phi}  {\bf 1}_{\widetilde{N}_c'}   &
0
\end{array}
\right), 
<M>  =
 \left(
\begin{array}{cc}
0  & 0 
 \\
0 & \Phi_0  {\bf 1}_{N_f'-\widetilde{N}_c'} 
\end{array}
\right)
\nonu \\
<Y> & = & <\widetilde{Y}> = <F> = <\widetilde{F}>= 0.
\label{vac}
\eea
Let us expand around on a point on (\ref{vac}), as done in
\cite{ISS}. 
Then the remaining relevant terms of superpotential are given by
\bea
W_{dual}^{rel} & = &  \Phi_0 \left( \delta \varphi  
\; \delta \widetilde{\varphi} + m \right) +
  \delta Z \; \delta \varphi  \; \widetilde{q}_0 
+ \delta \widetilde{Z} \; q_0  
\delta \widetilde{\varphi}
\nonu
\eea
by following the fluctuations for the various fields in \cite{Ahn07}.
Note that there exist three kinds of terms, 
the vacuum  $<q'>$ multiplied by 
$\delta Y \delta \widetilde{F}$,  
the vacuum  $<\widetilde{q'}>$ multiplied by $\delta F 
\delta \widetilde{Y}$, and 
the vacuum  $<\Phi>$ multiplied by $\delta Y 
\delta \widetilde{Y}$.
By redefining these, they do not enter the 
contributions for the one loop result, up to quadratic order. 
As done in \cite{Shih}, one gets 
that $m_{\Phi_0}^2$ will contain $(\log 4 -1) > 0$ implying that these
are stable.

Let us recombine $\widetilde{N}_c'$ flavor D4-branes among the $N_f'$ flavor 
D4-branes with those connecting NS5'-brane(coming from
$NS5_{-\theta}$-brane) 
and NS5-brane(coming from $NS5_{\theta'}$-brane) and then push
them in the $+v$ direction from the magnetic brane configuration 
in Figure 3. Then their mirrors will move $-v$
direction due to the presence of O6-plane. There are no color
D4-branes connecting NS5'-brane and NS5-brane and there exist only 
$(N_f'-\widetilde{N}_c')$ flavor D4-branes connecting D6-branes and
NS5'-brane(and their mirrors) that are misaligned to the
$\widetilde{N}_c'$ flavor D4-branes.

The minimal energy supersymmetry breaking brane configuration 
is given by Figure 5. If we are detaching NS5-brane to the $x^7$ direction, 
then this brane configuration leads to the one discussed in \cite{Ahn07-3} with
opposite RR charge of O6-plane where the gauge group was $Sp(N_c) \times
SU(\widetilde{N}_c')$ with fundamentals for each gauge group and 
bifundamentals. Of course, the brane configuration in Figure 5 with opposite
O6-plane and without a middle NS5-brane is exactly the same as the 
brane configuration of \cite{Ahn07-3}.  

\begin{figure}[ht]
   \epsfxsize=5.0in 
\centerline{\epsffile{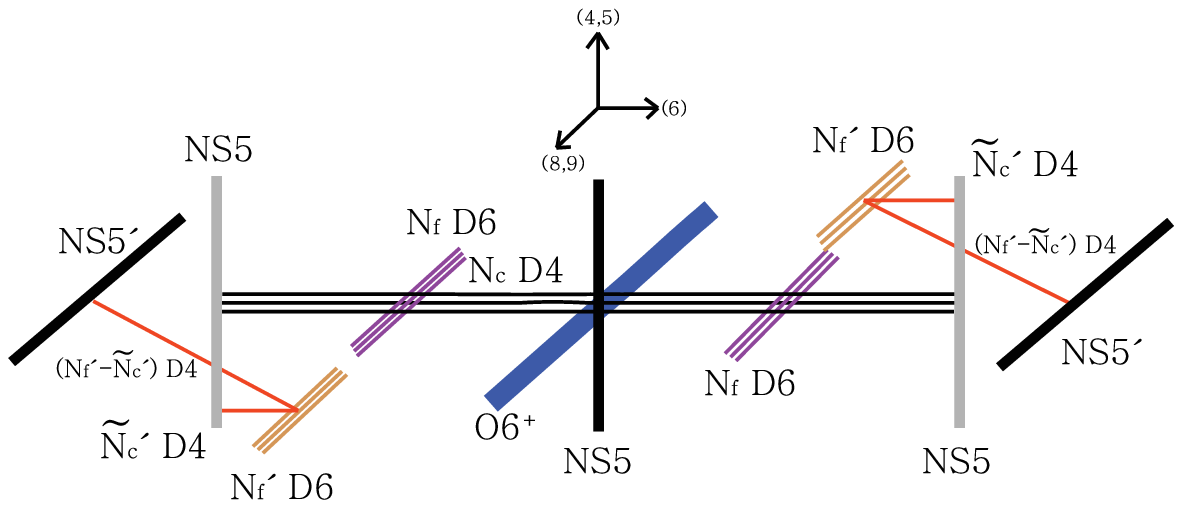}}
   \caption[FIG. \arabic{figure}.]{ 
The nonsupersymmetric minimal energy brane configuration of 
$SU(N_c) \times SU(\widetilde{N}_c'=N_f'+N_c-N_c')$ 
gauge group with fundamentals $Q,
\widetilde{Q}, q'$ and $\widetilde{q'}$ for each 
gauge group, the bifundamentals $Y$ and $\widetilde{Y}$ 
and 
a symmetric flavor $S$ and a
conjugate
symmetric flavor $\widetilde{S}$ for $SU(N_c)$
and various
gauge singlets.
We consider the massless case of $Q$ and $\widetilde{Q}$.
 Compared with Figure 3, there is a misalignment of the $N_f'$ flavor 
 D4-branes. Some of them are connecting to NS5'-brane and the other to
NS5-brane. In this figure, we put $\theta=\frac{\pi}{2}$ and $\theta'=0$.
The $N_c$ D4-branes can move freely along the $v$-diretion. }
\end{figure}

Since we are dealing with five NS5-branes, the magnetic M5-brane 
configuration with equal mass for $Q'$ and $\widetilde{Q'}$ and massless 
for $Q$ and $\widetilde{Q}$ 
can be characterized by the following quintic equation for $t$ 
as follows:
\bea
& & t^5 + \left[ v^{\widetilde{N}_c'} \right] t^4 + \left[ v^{N_c} \right]
 t^3 + \left[ (-1)^{N_c}
v^{N_c +N_f+ 2} 
(v +m')^{N_f'} \right] t^2 \nonu \\
& & + \left[ (-1)^{\widetilde{N}_c'+N_f+N_f'} v^{\widetilde{N}_c' +3N_f+6}  
(v+m')^{2N_f'} (v -m')^{N_f'} \right] t + 
\left[ v^{10+5N_f}  (v -m')^{2N_f'} 
 (v +m')^{3N_f'} \right] \nonu \\
& & = 0.
\nonu
\eea

Now the asymptotic regions of various NS5-branes 
can be determined by reading off the first two terms of the quintic
curve above
giving the
$NS5'_L$-brane
asymptotic region, next two terms giving 
$NS5_L$-brane asymptotic region, next two terms
giving $NS5_M$-brane asymptotic region, next two terms 
giving $NS5_R$-brane asymptotic region, and 
final two terms giving $NS5'_R$-brane asymptotic region.
Then the behavior of the supersymmetric M5-brane curves can be
summarized 
as follows:

1. $v \rightarrow \infty$ limit implies
\bea
w & \rightarrow & 0, \quad y \sim    v^{N_c-\widetilde{N}_c'} + \cdots \quad
\mbox{$NS5_{L}$ 
asymptotic region}, \nonu \\
w & \rightarrow  & 0, \quad y \sim    
v^{N_f+N_f'+2} + \cdots \quad
\mbox{$NS5_{M}$ asymptotic region}, 
\nonu \\
w & \rightarrow  & 0, \quad y \sim    
v^{\widetilde{N}_c-N_c+2N_f+2N_f'+4} + \cdots \quad
\mbox{$NS5_{R}$ asymptotic region}.   
\nonu
\eea

2.  $w \rightarrow \infty$ limit implies
\bea
v & \rightarrow &   -m', \quad 
y \sim  w^{\widetilde{N}_c'}
 +\cdots
\quad \mbox{$NS5_{L}'$ asymptotic region}, 
\nonu
\\
v & \rightarrow &  +m', \quad  
y \sim w^{-\widetilde{N}_c'+2N_f+2N_f'+4}
+\cdots
\quad \mbox{$NS5_{R}'$ asymptotic region}. 
\nonu
\eea

The two $NS5_{L,R}'$-branes 
are moving in the $\mp v$ direction holding everything 
else fixed instead of moving D6-branes in the $\mp v$ direction.
The corresponding mirrors of D4-branes are moved appropriately.
 Since the extra terms in the harmonic
function contribute to the $g(s)$ as a multiplication factor,   
the statement that $y$ is not equal to zero even if $v=0$,
implies that there exists the instability from a new M5-brane mode at
some point from the transition of SQCD-like theory description 
to M-theory description.

\section{Discussions}

So far, 
the intersecting brane configurations of
type 
IIA string theory are summarized by two figures Figure 4 and Figure 5
corresponding to the meta-stable nonsupersymmetric 
vacua of 
an ${\cal N}=1$ supersymmetric electric gauge theory 
with the gauge group $SU(N_c) \times SU(N_c')$ with fundamentals for each 
gauge group, the bifundamentals and a symmetric flavor and a
conjugate
symmetric flavor for $SU(N_c)$. This is done by applying Seiberg dual 
to  each gauge group  independently and obtaining two ${\cal N}=1$
supersymmetric 
dual magnetic gauge theories with dual matters including the gauge
singlets. 

One can also 
generalize to the same $SU(N_c) \times SU(N_c')$ gauge theory
with
fundamentals for each 
gauge group, the bifundamentals and an antisymmetric flavor and a
conjugate
symmetric flavor for $SU(N_c)$. 
Let us describe how the intersecting brane configurations arise here. 

\subsection{ Electric theory with $SU(N_c) \times SU(N_c')$ gauge group}

The gauge group is $SU(N_c) \times SU(N_c')$ which is the same as
before but the matter contents
are different and are given as follows:

$\bullet$
$N_f$-chiral multiplets $Q$ are  in the
representation $({\bf N_c,1
})$, and 
$N_f$-chiral multiplets $\widetilde{Q}$ are in  
the representation $({\bf \overline{N_c}, 1})$,
under the gauge group

$\bullet$
Eight-chiral multiplets $\hat{Q}$ are  in the
representation $({\bf N_c,1})$
under the gauge group

$\bullet$
$N_f'$-chiral multiplets $Q'$ are  in the
representation $({\bf 1, N_c'
})$, and 
$N_f'$-chiral multiplets $\widetilde{Q'}$ are in  
the representation $({\bf 1,\overline{N_c'}})$,
under the gauge group

$\bullet$
The flavor-singlet field $X$ is 
in the bifundamental representation $({\bf N_c, \overline{N_c'} })$, 
and its conjugate field $\widetilde{X}$
 is 
in the bifundamental representation $({\bf \overline{N_c}, N_c'})$, 
under the gauge group

$\bullet$ The flavor-singlet field $A$, which is 
in an antisymmetric tensor representation under the $SU(N_c)$, is in the
representation $({\bf \frac{1}{2} N_c(N_c-1),1})$, and
its conjugate field $\widetilde{S}$ is in the 
representation $({\bf \overline{\frac{1}{2} N_c(N_c+1)},1})$, under the
gauge group

If there are no antisymmetric, conjugate symmetric tensors and eight
fundamentals, 
this theory 
is described by the work of \cite{ILS,BH,BIWW} from field theory analysis or
corresponding brane
dynamics.  Ignoring the presence of the fields $Q', \widetilde{Q'}, X$ and
$\widetilde{X}$,
then this theory will reduce to a single gauge group $SU(N_c)$ with an 
antisymmetric flavor,
conjugate symmetric flavor and fundamental flavors $A, \widetilde{S},
Q$ and $\widetilde{Q}$ studied in  
\cite{ILS,LLL1,BHKL,EGKT,Ahn07-1}.

The coefficient of the beta function of the first gauge group 
is given by
\bea
b_{SU(N_c)}=3N_c-(N_f+4)-N_c'-\frac{1}{2}(N_c+2)-\frac{1}{2}(N_c-2)
\nonu
\eea
by realizing the index of the antisymmetric and symmetric
representations
of $SU(N_c)$ gauge group
and 
the coefficient of the beta function of the second gauge group  
is given by
\bea
b_{SU(N_c')}=3N_c'-N_f'-N_c.
\nonu
\eea
These values will change when we go to the magnetic theory.
The classical superpotential is given by
\bea
W & = & \mu A_d^2 + A_d A \widetilde{S} + \lambda Q A_d \widetilde{Q}
+ \mu' A_d^{'2} + \lambda' Q' A_d' \widetilde{Q'} + X A_d \widetilde{X} + 
\widetilde{X} A_d' X \nonu \\
&+& \hat{Q} \widetilde{S} \hat{Q}
+ m Q \widetilde{Q} + m' Q' \widetilde{Q'},
\label{pot}
\eea
where the coefficient functions are given by
\bea
\qquad \mu \equiv \tan (\frac{\pi}{2} -\theta), 
\qquad \mu' \equiv \tan (\theta'-\theta), \qquad
        \lambda \equiv \sin (\theta-\omega), \qquad \lambda' \equiv
        \sin 
(\theta'-\theta-\omega'). 
\nonu
\eea
Here the adjoint field for $SU(N_c)$ gauge group is denoted by $A_d$ while
the adjoint field for $SU(N_c')$ gauge group is denoted by $A_d'$.
The second term in (\ref{pot}) arises 
from the presence of antisymmetric flavor $A$ and a conjugate
symmetric flavor $\widetilde{S}$. Except this, the
eighth term and the last
mass terms, the above superpotential becomes 
the one described in \cite{BHKL,BH}.
Setting the fields $Q', \widetilde{Q'}, X, \widetilde{X}$ and $A_d'$ to zero, 
the superpotential becomes the one discussed in \cite{LLL1,BHKL,EGKT,Ahn07-1}.
After integrating out the adjoint fields $A_d$ and $A_d'$, 
this superpotential at the particular rotation angles $\theta=0$ and 
$\theta'=\frac{\pi}{2}$ 
will reduce to the last two mass-deformed terms since the coefficient functions 
$\frac{1}{\mu}$
and $\frac{1}{\mu'}$ vanish at this particular rotation angles.
For the nonsupersymmetric brane configuration, we will
consider this particular brane configuration with the constraint 
$\theta=0$ and 
$\theta'=\frac{\pi}{2}$ all the time.

The type IIA brane configuration for this gauge theory can be
constructed similarly and can be drawn as Figure 1 except that at the
origin of $(x^6, v,w)$ coordinates, there exist NS5'-brane,
$O6^{+}$-plane, $O6^{-}$-plane and eight half D6-branes, instead of
having NS5-brane and $O6^{+}$-plane.  One can denote this as 
$NS5'/O6/D6$-branes, as described in \cite{Ahn07-1}.
If we are detaching $NS5_{\pm \theta'}$-branes, $D6_{\pm \omega'}$-branes
and $N_c'$ $D4$-branes(and its mirrors), this brane configuration will 
reduce to the one described in \cite{LLL1,BHKL,EGKT,Ahn07-1}.
If we are detaching a middle NS5-brane to the $x^7$ direction, 
then this will lead to the brane configuration \cite{LO,Ahn07-3} 
with the gauge group 
$Sp(N_c) \times SU(N_c')$ or $SO(N_c) \times SU(N_c')$ 
with fundamentals for 
each gauge group and bifundamentals, depending on the RR charge of O6-plane. 
If we are detaching all the branes living on the negative $x^6$ region,
two O6-planes and eight D6-branes, 
then this will become the brane configuration of \cite{BH}, as we
mentioned before.

\subsection{ Magnetic theory with 
$SU(\widetilde{N}_c) \times SU(N_c')$ gauge group}

From the magnetic and electric brane configurations we did not present
here,  
the linking number counting, as done in \cite{Ahn07-1}, implies that 
the number of dual color $\widetilde{N}_c$ is given by
\bea
\widetilde{N}_c = 2(N_f+N_c')-N_c+4.
\label{Dualnum}
\eea
Compared with the electric theory, the magnetic brane configuration
has 
 different features where there are
extra $N_f$ D6-branes and newly created $N_f$ flavor D4-branes for the
gauge group $SU(N_c')$.

The dual magnetic gauge group is  
$SU(\widetilde{N}_c) \times SU(N_c')$ 
and the matter contents 
are as follows:

$\bullet$ 
$N_f$-chiral multiplets $q$ are  in the
representation $({\bf \widetilde{N}_c}, 1)$,
$N_f$-chiral multiplets $\widetilde{q}$ are in the representation 
$({\bf \overline{\widetilde{N}_c}}, 1)$,
under the gauge group

$\bullet$ 
Eight-chiral multiplets $\hat{q}$ are  in the
representation $({\bf \widetilde{N}_c}, 1)$
under the gauge group

$\bullet$
$N_f'$-chiral multiplets $Q'$ are  in the
representation $({\bf 1, N_c'
})$, and 
$N_f'$-chiral multiplets $\widetilde{Q'}$ are in  
the representation $({\bf 1,\overline{N_c'}})$,
under the gauge group

$\bullet$
The flavor-singlet field $Y$ is 
in the bifundamental representation $({\bf \widetilde{N}_c, 
\overline{N_c'} })$, 
and its complex conjugate field $\widetilde{Y}$
 is 
in the bifundamental representation $({\bf \overline{\widetilde{N}_c}, 
N_c'})$, 
under the gauge group

$\bullet$ The flavor-singlet field $a$, which is 
in an antisymmetric tensor representation 
under the $SU(\widetilde{N}_c)$, is in the
representation $({\bf \frac{1}{2} \widetilde{N}_c(\widetilde{N}_c-1),1})$, and
the conjugate symmetric field $\widetilde{s}$ is in the 
representation $({\bf \overline{\frac{1}{2} 
\widetilde{N}_c(\widetilde{N}_c+1)},1})$, under the
gauge group

There are also $(N_f+N_c')^2$ gauge-singlets in the first dual gauge group
factor
as follows:

$\bullet$
$N_f$-fields $F'$ are  in the representation $({\bf 1, N_c' })$, 
and its complex conjugate
$N_f$-fields $\widetilde{F'}$ are in the representation 
$({\bf 1, \overline{N_c'} })$, 
under the gauge group

$\bullet$
$N_f^{2}$-fields $M'$ are in the representation $({\bf 1, 1})$ under the
gauge group

$\bullet$
The $N_c^{'2}$-fields 
$\Phi'$ is in the representation $({\bf 1, N_c^{'2}-1}) \oplus ({\bf 1,1
})$ 
under the gauge group  

Moreover, there are additional $N_f(2N_f+1)$ gauge singlets

$\bullet$
$N_f^{2}$-fields $N'$ are in the representation $({\bf 1, 1})$ under the
gauge group

$\bullet$
$N_f$-fields $\widetilde{M}$ are in the representation $({\bf 1, 1})$ 
under the
gauge group

$\bullet$ $\frac{1}{2} N_f(N_f+1)$-fields $P'$ are 
 in the representation $({\bf 1, 1})$, and its 
conjugate $\frac{1}{2} N_f(N_f+1)$-fields 
 $\widetilde{P'}$  are 
 in the representation $({\bf 1, 1})$,
under the
gauge group

These are represented by $N' \equiv Q \widetilde{S} A
\widetilde{Q}, \widetilde{M} \equiv \hat{Q} \widetilde{Q},
P' \equiv Q \widetilde{S} Q$ and $\widetilde{P'} \equiv
\widetilde{Q} A \widetilde{Q}$ in terms of fields in electric theory, 
as observed in \cite{ILS,Ahn07-1}. 
Although these gauge-singlets $N', P'$, and $\widetilde{P'}$ 
appear in the dual magnetic
superpotential for the general rotation angles $\theta$ and $\theta'$,
the case $\theta=0$ we are considering 
does not contain these gauge singlets, as found in \cite{Ahn07-1}. 

The type IIA magnetic brane configuration can be described as in
Figure 2 similarly, by
replacing NS5-brane and $O6^{+}$-plane by the combination of 
$NS5'/O6/D6$-branes around at the origin, as mentioned before. 

The coefficient of the beta function of the first dual gauge group factor 
is given by
\bea
b_{SU(\widetilde{N}_c)}^{mag}=3\widetilde{N}_c-(N_f+4)-N_c'-
\frac{1}{2}(\widetilde{N}_c+2)-\frac{1}{2}(\widetilde{N}_c-2)
\nonu
\eea
and 
the coefficient of the beta function of the second gauge group factor 
is given by
\bea
b_{SU(N_c')}^{mag}=3N_c'-N_f'-\widetilde{N}_c-N_f-N_c'.
\nonu
\eea
It is clear
that the $SU(N_c')$ fields in the magnetic theory 
are different from those of the electric theory and this will lead to
the various interaction terms in the dual magnetic superpotential
\footnote{More explicitly the conditions
  $b_{SU(\widetilde{N}_c)}^{mag} < 0$ and $b_{SU(N_c)} > 0$ imply
that $N_f+N_c' < \frac{2}{3} N_c -\frac{4}{3}$. Also the number of
dual colors $\widetilde{N}_c$ defined as (\ref{Dualnum}) 
should be positive. Then the range for the $N_f$ in the first gauge
group can be written as $ \frac{1}{2} N_c-2 < N_f+N_c' 
< \frac{2}{3} N_c -\frac{4}{3}$. 
Since $b_{SU(N_c')}-b_{SU(N_c')}^{mag} = 3(N_c'+N_f)-2N_c +4 < 0$,
if we require that $b_{SU(N_c')}^{mag} < 0$ which is equivalent to 
$N_c-3N_f-4 < N_f'$, then the electric description of $SU(N_c')$ is IR
free because $b_{SU(N_c')} < 0$.
At high energy, $SU(N_c')$ theory is strongly coupled while $SU(N_c)$
theory
is UV free. At the scale $\Lambda_1$, the $SU(N_c)$ theory is strongly
coupled and the Seiberg duality occurs. All the running couplings are
changed by this duality and all the coefficients of beta functions, 
$b_{SU(\widetilde{N}_c)}^{mag}$ and $b_{SU(N_c')}^{mag}$
become negative. Then at energy scale lower than $\Lambda_1$, 
the theory is weakly coupled. When 
$b_{SU(N_c')}^{mag} < b_{SU(\widetilde{N}_c)}^{mag} < 0$, 
the one loop computation is reliable with $\Lambda_1 << \Lambda_2$. 
When 
$b_{SU(\widetilde{N}_c)}^{mag} < b_{SU(N_c')}^{mag} < 0$, 
the requirement that 
$SU(N_c')^{mag}$ theory is less coupled than the 
$SU(\widetilde{N}_c)^{mag}$ at the supersymmetry breaking scale $\mu$
provides a stronger constraint on $\Lambda_2$. 
Then under the constraint, $\Lambda_2 >> \left(
  \frac{\Lambda_1}{\mu}\right)^b
\Lambda_1$
where $b$ is defined as in footnote 1, 
one can ignore the contribution from the gauge coupling of 
$SU(N_c')^{mag}$ at the supersymmetry breaking scale and one relies on
the one loop computation. 
}.

The dual magnetic superpotential 
\footnote{
 We integrate out the bifundamentals $Y$ and $\widetilde{Y}$ and $\hat{q}$
in such a way that the gauge group $SU(\widetilde{N}_c)$ is not broken 
by the fields $Y$ and $\widetilde{Y}$ and $\hat{q}$, as in meta-stable state, so 
$< Y > = 0 = < \widetilde{Y} >=< \hat{q} >$ in (\ref{vacuum}). 
For nonzero vacuum expectation values for $M'$, this
superpotential gives the $SU(\widetilde{N}_c)$  ``flavors''
$q \widetilde{s}$ and $a \widetilde{q}$, the mass $<M'>$. Below the energy scale 
$<M'>$, one can integrate out these massive flavors using the equations
of motion $< q \widetilde{s} > = 0 = < a \widetilde{q} >$. 
Then the low energy theory has a superpotential term 
 which is proportional to $\left( 
\widetilde{\Lambda}^{2\widetilde{N}_c-N_f-N_c'-4} \mbox{det} M' 
\right)^{\frac{1}{\widetilde{N}_c}}$ plus $m M'$. There is no
conserved $U(1)_R$ symmetry because it is anomalous under the gauged 
$SU(\widetilde{N}_c)$ in the sense that the determinant term above
breaks it explicitly. 
Therefore, the $U(1)_R$ symmetry returns
an ``approximate'' accidental symmetry of the IR theory.  }
for massless fundamental flavors
 $Q'$
and $\widetilde{Q'}$ and massive fundamental flavors 
$Q$ and $\widetilde{Q}$ is given by
\bea
W_{dual}= \left(M' q \widetilde{s} a \widetilde{q} +  m M' + \hat{q}
 \widetilde{s} \hat{q} +\widetilde{M} \hat{q} \widetilde{q} \right)+ 
\widetilde{Y} \widetilde{F'} q +
Y \widetilde{q} F' + \Phi' Y \widetilde{Y} +\left( \Phi^{'2} + \cdots \right)
\label{ppot}
\eea
where the mesons are given in terms of fields in the electric theory 
\bea
M' \equiv Q \widetilde{Q}, \qquad 
 \widetilde{M} \equiv \hat{Q} \widetilde{Q}, \qquad 
F' \equiv \widetilde{X} Q, \qquad 
\widetilde{F'} \equiv X \widetilde{Q}, \qquad
\Phi' \equiv X \widetilde{X}.
\nonu
\eea
The last piece $\left( \Phi^{'2} + \cdots \right)$ in $W_{dual}$
is coming from the superpotential (\ref{pot}) 
in an electric theory and  contains also $ N' q \widetilde{q} + P' q 
\widetilde{s} q +\widetilde{P'} \widetilde{q} a \widetilde{q}$ for
 the general rotation angles $\theta, \theta', \omega$ and $\omega'$.  
When $\theta=0$ and $\theta'=\frac{\pi}{2}$, this will vanish and the
superpotential consists of the first seven terms in (\ref{ppot}) which
will play an important role when we discuss 
the nonsupersymmetric brane configuration.
Also the above mesons can be interpreted as strings 
connecting various D-branes, as we did in previous section.

The dual quarks $q$ and $\widetilde{q}$ are fundamental 
$({\bf \widetilde{N}_c,1})$ and
antifundamental $({\bf \overline{\widetilde{N}_c}},1)$ 
for the gauge group indices and antifundamentals for
the flavor indices. 
The flavor-singlet fields $a$ and $\widetilde{s}$ are antisymmetric  
$({\bf \frac{1}{2} \widetilde{N}_c(\widetilde{N}_c-1),1})$ and conjugate
symmetric tensor 
 $({\bf \overline{\frac{1}{2} 
\widetilde{N}_c(\widetilde{N}_c+1)},1})$
for the gauge group indices respectively. The quantity $q
\widetilde{s} a \widetilde{q}$ has a rank $\widetilde{N}_c$ and the
mass matrix $m$ has a rank $N_f$. Then the F-term equation for $M'$
cannot be satisfied if the rank $N_f$ exceeds $\widetilde{N}_c$ and
the supersymmetry is broken.

The classical moduli space of vacua can be obtained from F-term
equations and one gets
\bea
 q \widetilde{s} a \widetilde{q} +  m & = & 0, \qquad
\widetilde{s} a \widetilde{q} M' + \widetilde{Y} \widetilde{F}   =  0, \nonu \\
a \widetilde{q} M' q  + \hat{q} \hat{q} & = & 0, \qquad
\widetilde{q} M' q \widetilde{s}  =  0, \nonu \\
  M' q \widetilde{s} a + \widetilde{M} \hat{q} + F' Y & = & 0, \qquad
\widetilde{F'} q + \Phi' Y  =  0,
\nonu \\
q \widetilde{Y} & = & 0, \qquad
\widetilde{q} F' + \widetilde{Y} \Phi'  =  0, \nonu \\
Y \widetilde{q} & = & 0, \qquad
Y \widetilde{Y} + \Phi'  =  0, 
\nonu \\
\widetilde{s} \hat{q} + \widetilde{q} \widetilde{M} & = & 0, \qquad
\hat{q} \widetilde{q} =0.
\nonu
\eea
Some of F-term equations are satisfied if one takes the zero vacuum
expectation values for the fields $Y, \widetilde{Y}, F', 
\widetilde{F'}, \hat{q}$ and $\widetilde{M}$. 
Then, it is easy to see that 
\bea
a \widetilde{q} M' =0= M' q \widetilde{s}, \qquad
 q \widetilde{s} a \widetilde{q} +  m  =  0.
\nonu 
\eea
Then the solutions can be written as
\bea
<q \widetilde{s}>  & = &  \left(
\begin{array}{c}
\sqrt{m} e^{\phi} {\bf 1}_{\widetilde{N}_c}  \\
0
\end{array}
\right),  
<a \widetilde{q}> =
 \left(
\begin{array}{cc}
\sqrt{m} e^{-\phi}  {\bf 1}_{\widetilde{N}_c}   &
0
\end{array}
\right), 
<M'>  =
 \left(
\begin{array}{cc}
0  & 0 
 \\
0 & \Phi_0  {\bf 1}_{N_f-\widetilde{N}_c} 
\end{array}
\right)
\nonu \\
<Y> & = & <\widetilde{Y}> = <F'> = <\widetilde{F'}>=<\hat{q}> = 
<\widetilde{M}>= 0.
\label{vacuum}
\eea
Let us expand around on a point on (\ref{vacuum}), as done in
\cite{ISS}. 
Then the remaining relevant terms of superpotential are given by
\bea
W_{dual}^{rel} & = &  \Phi_0 \left( \delta \hat{\varphi}  
\; \delta \hat{\widetilde{\varphi}} + m \right) +
  \delta Z \; \delta \hat{\varphi} \; a_0 \; \widetilde{q}_0 
+ \delta \widetilde{Z} \; q_0 \; \widetilde{s}_0 \;
\delta \hat{\widetilde{\varphi}}
\nonu
\eea
by following the fluctuations for the various fields in \cite{Ahn07-1}.
Note that there exist five kinds of terms, 
the vacuum  $<q>$ multiplied by 
$\delta \widetilde{Y} \delta \widetilde{F'}$,  
the vacuum  $<\widetilde{q}>$ multiplied by $\delta F' 
\delta Y$, 
the vacuum  $<\Phi'>$ multiplied by $\delta Y 
\delta \widetilde{Y}$, the vacuum $<\widetilde{s}>$ multiplied by 
$\delta \hat{q} \delta \hat{q}$, and the vacuum $\widetilde{q}$
multiplied by $\delta \widetilde{M} \delta \hat{q}$.
By redefining these as before, they do not enter the 
contributions for the one loop result, up to quadratic order. 
As done in \cite{Ahn07}, the defining function ${\cal F}(v^2)$ can be
computed
and using the equation (2.14) of \cite{Shih} 
of $m_{\Phi_0}^2$ and ${\cal F}(v^2)$, one gets 
that $m_{\Phi_0}^2$ will contain $(\log 4 -1) > 0$ implying that these
are stable.

Then the minimal energy supersymmetry breaking brane configuration 
is given by Figure 6. If we are detaching $N_c'$ D4-branes, $N_f'$
D6-branes and NS5'-brane(coming from $NS5_{\theta'}$-brane)(and its
mirrors), 
then this brane configuration leads to the one described in \cite{Ahn07-1}.

\begin{figure}[ht]
   \epsfxsize=5.0in 
\centerline{\epsffile{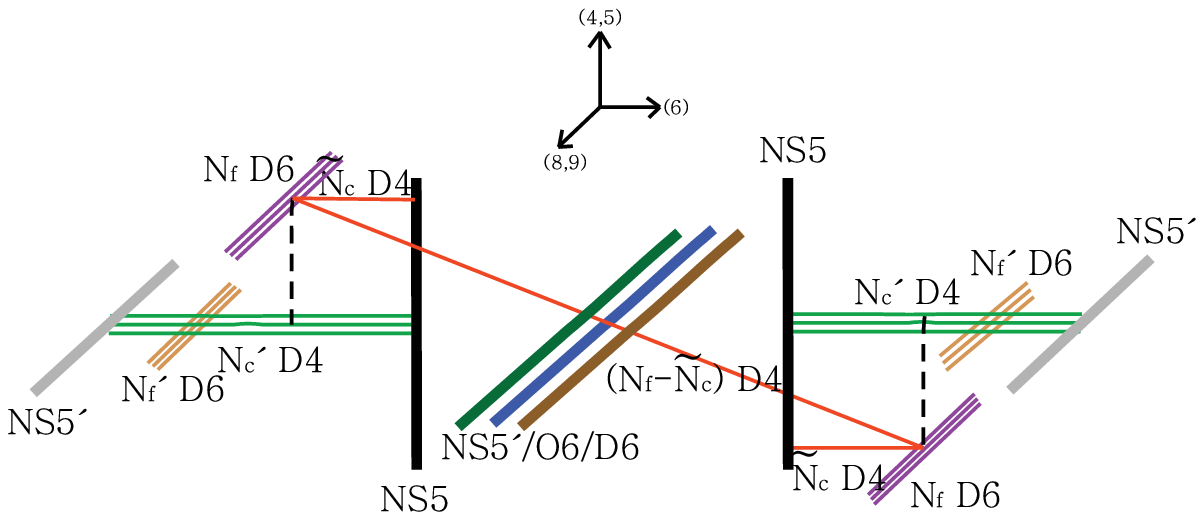}}
   \caption[FIG. \arabic{figure}.]{ 
The nonsupersymmetric minimal energy brane configuration of 
$SU(\widetilde{N}_c=2N_f +2N_c'-N_c+4) \times SU(N_c')$ 
 gauge group with fundamentals $q,
\widetilde{q}, \hat{q}, Q'$ and $\widetilde{Q'}$ for each 
gauge group, the bifundamentals $Y$ and $\widetilde{Y}$ and 
an antisymmetric flavor $a$ and a
conjugate
symmetric flavor $\widetilde{s}$ for $SU(\widetilde{N}_c)$ and various
gauge singlets. 
We consider the massless case of $Q'$ and $\widetilde{Q'}$.
From the modified Figure 2, there is a mislaignment of $N_f$ flavor
D4-branes.
Some of them are connecting to NS5'-brane and the other to
NS5-brane. In this figure, we put $\theta=0$ and $\theta'=\frac{\pi}{2}$.  }
\end{figure}

Now the asymptotic regions of various NS5-branes, at nonzero string
coupling constant, 
can be determined by reading off the first two terms of the quintic
curve given in subsection 3.1
giving the
$NS5'_L$-brane
asymptotic region, next two terms giving 
$NS5_L$-brane asymptotic region, next two terms
giving $NS5'_M$-brane asymptotic region, next two terms 
giving $NS5_R$-brane asymptotic region, and 
final two terms giving $NS5'_R$-brane asymptotic region.
Then the behavior of the supersymmetric M5-brane curves can be
summarized 
as follows:

1. $v \rightarrow \infty$ limit implies
\bea
w & \rightarrow & 0, \quad y \sim    
v^{\widetilde{N}_c-N_c'+N_f+N_f'} + \cdots \quad
\mbox{$NS5_{L}$ 
asymptotic region}, \nonu \\
w & \rightarrow  & 0, \quad y \sim    
v^{N_c'-\widetilde{N}_c+N_f+N_f'+4} + \cdots \quad
\mbox{$NS5_{R}$ asymptotic region}.   
\nonu
\eea

2.  $w \rightarrow \infty$ limit implies
\bea
v & \rightarrow &   0, \quad 
y \sim  w^{N_c'}
 +\cdots
\quad \mbox{$NS5_{L}'$ asymptotic region}, 
\nonu
\\
v & \rightarrow &   \pm m, \quad 
y \sim  w^{N_f+N_f'+2}
 +\cdots
\quad \mbox{$NS5_{M}'$ asymptotic region}, 
\nonu
\\
v & \rightarrow &  0, \quad  
y \sim w^{2N_f+2N_f'-N_c'+4}
+\cdots
\quad \mbox{$NS5_{R}'$ asymptotic region}. 
\nonu
\eea

The harmonic function in the Tau-NUT space, sourced by $2N_f$
D6-branes, O6-plane and $2N_f'$ D6-branes, can be determined once we
fix the $x^6$ position for these branes and write the charges. 
Then the first order
differential equation for the $g(s)$ where the absolute value of
$g(s)$ is equal to the absolute value of $w$ can be solved exactly with the
appropriate boundary conditions on $NS5_M'$ asymptotic
region from above classification 2. Since the extra terms in the harmonic
function contribute to the $g(s)$ as a multiplication factor,   
the contradiction with the correct statement that $y$ should 
vanish only if $v=0$,
implies that there exists the instability from a new M5-brane mode.

\subsection{ Magnetic theory with 
$SU(N_c) \times SU(\widetilde{N}_c')$ gauge group}

Now we continue to analyze for other magnetic theory.
From the magnetic and electric brane configurations,  
the linking number counting implies that 
the number of dual color $\widetilde{N}_c'$ is given by, 
as we did in (\ref{num}),
\bea
\widetilde{N}_c' = N_f' + N_c -N_c'.
\label{Num}
\eea

The dual magnetic gauge group is  
given by $SU(N_c) \times SU(\widetilde{N}_c')$ 
and the matter contents 
are as follows:

$\bullet$
$N_f$-chiral multiplets $Q$ are  in the
representation $({\bf N_c,1
})$, and 
$N_f$-chiral multiplets $\widetilde{Q}$ are in  
the representation $({\bf \overline{N_c}, 1})$,
under the gauge group

$\bullet$
Eight-chiral multiplets $\hat{Q}$ are  in the
representation $({\bf N_c,1
})$
under the gauge group

$\bullet$ 
$N_f'$-chiral multiplets $q'$ are  in the
representation $({\bf 1,\widetilde{N}_c'})$,
$N_f'$-chiral multiplets $\widetilde{q'}$ are in the representation 
$({\bf 1, \overline{\widetilde{N}_c'}})$,
under the gauge group

$\bullet$
The flavor singlet field $Y$ is 
in the bifundamental representation $({\bf N_c, 
\overline{\widetilde{N}_c'} })$, 
and its complex conjugate field $\widetilde{Y}$
 is 
in the bifundamental representation $({\bf \overline{N_c}, 
\widetilde{N}_c'})$, 
under the gauge group

$\bullet$ The flavor singlet field $A$, which is 
in an  antisymmetric tensor representation under the $SU(N_c)$, is in the
representation $({\bf \frac{1}{2} N_c(N_c-1),1})$, and
its conjugate field $\widetilde{S}$ is in the 
representation $({\bf \overline{\frac{1}{2} N_c(N_c+1)},1})$, under the
gauge group

There are $(N_f'+N_c)^2$ gauge singlets in the second dual gauge group
factor
as follows:

$\bullet$
$N_f'$-fields $F$ are  in the representation $({\bf N_c,1 })$, 
and its complex conjugate
$N_f'$-fields $\widetilde{F}$ are in the representation 
$({\bf \overline{N_c},1 })$, 
under the gauge group

$\bullet$
$N_f^{'2}$-fields $M$ are in the representation $({\bf 1, 1})$ under the
gauge group

$\bullet$
The $N_c^2$-fields 
$\Phi$ is in the representation $({\bf N_c^2-1, 1}) \oplus ({\bf 1,1
})$ 
under the gauge group  

The type IIA magnetic brane configuration can be described as in Figure 3, by
replacing NS5-brane and $O6^{+}$-plane by the combination of 
$NS5'/O6/D6$-branes around at the origin, as mentioned before. 
We'll not present those here.

The coefficient of the beta function of the first gauge group factor 
is given by
\bea
b_{SU(N_c)}^{mag}=3N_c-(N_f+4)-\widetilde{N}_c'-N_f'-N_c-
\frac{1}{2}(N_c+2)-\frac{1}{2}(N_c-2)
\nonu
\eea
as before and 
the coefficient of the beta function of the second gauge group factor 
is given by
\bea
b_{SU(\widetilde{N}_c')}^{mag}=3\widetilde{N}_c'-N_f'-N_c.
\nonu
\eea
Since $b_{SU(N_c)}-b_{SU(N_c)}^{mag} > 0$, the $SU(N_c)$ theory is more
asymptotically free 
 \footnote{More explicitly, the conditions
  $b_{SU(\widetilde{N}_c')}^{mag} < 0$ and $b_{SU(N_c')} > 0$ imply
that $N_f'+N_c < \frac{3}{2} N_c' $. Also the number of
dual colors $\widetilde{N}_c'$ defined as (\ref{Num}) 
should be positive. Then the range for the $N_f'$ in the second gauge
group can be written as $ N_c' < N_f'+N_c 
< \frac{3}{2} N_c'$. 
 The condition
$b_{SU(N_c)}^{mag} < 0$ implies 
 $ N_c'-2N_f'-4 < N_f$.
The $b_{SU(N_c)}$ can be IR free or UV free in the electric description. 
One can easily analyze four different possibilities as in footnote 3.
Two cases for the positivity or negativity of the difference between 
$b_{SU(\widetilde{N}_c')}^{mag}$ and $b_{SU(N_c)}^{mag}$ and 
two cases for the UV free or IR free for $SU(N_c)$ theory in an
electric description.
}
than the $SU(N_c)^{mag}$ theory \cite{BIWW}.

The dual magnetic superpotential 
\footnote{
For nonzero vacuum expectation values for $M$, this
superpotential gives the $SU(\widetilde{N}_c')$ fundamental flavors
$q'$ and $\widetilde{q'}$, the mass $<M>$. Below the energy scale 
$<M>$, one can integrate out these massive flavors using the equations
of motion $< q' > = 0 = < \widetilde{q'} >$. 
Then the low energy theory has a superpotential term 
 which is proportional to $\left( 
\widetilde{\Lambda}^{3\widetilde{N}_c'-N_f'-N_c} \mbox{det} M 
\right)^{\frac{1}{\widetilde{N}_c'}}$ plus $m' M$. There is no
conserved $U(1)_R$ symmetry because it is anomalous under the gauged 
$SU(\widetilde{N}_c')$ in the sense that the determinant term above
breaks it explicitly. Remember that the $R$ charges for the fields are
as follows: $R(Y)=R(q')=\frac{N_c'}{N_f'+N_c}$ and $R(M)=R(F)=R(\Phi)=
2-\frac{2N_c'}{N_f'+N_c}$. Therefore, the $U(1)_R$ symmetry returns an
 ``approximate'' accidental symmetry of the IR theory.  }
for massless fundamental flavors
$Q$
and $\widetilde{Q}$(i.e., $m=0$) and massive fundamental flavors $Q'$
and $\widetilde{Q'}$ is 
\bea
W_{dual}'= \left( M q' \widetilde{q'}  + m' M \right) + Y \widetilde{F} q' +
\widetilde{Y} \widetilde{q'} F + \Phi Y \widetilde{Y} 
+ \left( \Phi^2 + \cdots \right)
\label{Dual}
\eea
where  the mesons are given by
\bea
M \equiv Q' \widetilde{Q'}, \qquad 
F \equiv X Q', \qquad 
\widetilde{F} \equiv \widetilde{X} \widetilde{Q'}, \qquad
\Phi \equiv X \widetilde{X}.
\nonu
\eea
Here the last piece $\left( \Phi^2 + \cdots \right)$ in $W_{dual}$
is coming from the superpotential (\ref{pot}) 
in an electric theory for
 the general rotation angles $\theta, \theta', \omega$ and $\omega'$.  
When $\theta=0$ and $\theta'=\frac{\pi}{2}$, this will vanish and the
superpotential consists of the first five terms in (\ref{Dual}).
Also the above mesons can be interpreted as strings 
connecting various D-branes, as before.

The dual quarks $q'$ and $\widetilde{q'}$ are 
fundamental $({\bf 1, \widetilde{N}_c'})$ and
antifundamental $({\bf 1, \overline{\widetilde{N}_c'}})$ 
for the gauge group indices and antifundamentals for
the flavor indices. 
The quantity $q'
 \widetilde{q'}$ has a rank $\widetilde{N}_c'$ and the
mass matrix $m'$ has a rank $N_f'$. Then the F-term equation for $M$
cannot be satisfied if the rank $N_f'$ exceeds $\widetilde{N}_c'$ and
the supersymmetry is broken.

The classical moduli space of vacua can be obtained from F-term
equations and one gets
\bea
 q'  \widetilde{q'} +  m' & = & 0, \qquad
\widetilde{q'} M + Y \widetilde{F}   =  0, \nonu \\
  M q'  + F \widetilde{Y} & = & 0, \qquad
\widetilde{F} q' +  \widetilde{Y} \Phi  =  0,
\nonu \\
q' Y & = & 0, \qquad
\widetilde{q'} F +  \Phi Y  =  0, \nonu \\
\widetilde{Y} \widetilde{q'} & = & 0, \qquad
Y \widetilde{Y}  =  0. 
\nonu
\eea
Other F-term equations are satisfied if one takes the zero vacuum
expectation values for the fields $Y, \widetilde{Y}, F$ and 
$\widetilde{F}$.
Then, it is easy to see that 
\bea
\widetilde{q'} M =0= M q', \qquad
 q'  \widetilde{q'} +  m'  =  0.
\nonu 
\eea
Then the solutions can be written as
\bea
<q' >  & = &  \left(
\begin{array}{c}
\sqrt{m} e^{\phi} {\bf 1}_{\widetilde{N}_c'}  \\
0
\end{array}
\right),  
< \widetilde{q'}> =
 \left(
\begin{array}{cc}
\sqrt{m} e^{-\phi}  {\bf 1}_{\widetilde{N}_c'}   &
0
\end{array}
\right), 
<M>  =
 \left(
\begin{array}{cc}
0  & 0 
 \\
0 & \Phi_0  {\bf 1}_{N_f'-\widetilde{N}_c'} 
\end{array}
\right)
\nonu \\
<Y> & = & <\widetilde{Y}> = <F> = <\widetilde{F}>= 0.
\label{vac1}
\eea
Let us expand around on a point on (\ref{vac1}), as done in
\cite{ISS}. 
Then the remaining relevant terms of superpotential are given by
\bea
W_{dual}^{rel} & = &  \Phi_0 \left( \delta \varphi  
\; \delta \widetilde{\varphi} + m \right) +
  \delta Z \; \delta \varphi  \; \widetilde{q}_0 
+ \delta \widetilde{Z} \; q_0  
\delta \widetilde{\varphi}
\nonu
\eea
by following the fluctuations for the various fields in \cite{Ahn07}.
Note that there exist three kinds of terms, 
the vacuum  $<q'>$ multiplied by 
$\delta Y \delta \widetilde{F}$,  
the vacuum  $<\widetilde{q'}>$ multiplied by $\delta F 
\delta \widetilde{Y}$, and 
the vacuum  $<\Phi>$ multiplied by $\delta Y 
\delta \widetilde{Y}$.
By redefining these, they do not enter the 
contributions for the one loop result, up to quadratic order. 
As done in \cite{Shih}, one gets 
that $m_{\Phi_0}^2$ will contain $(\log 4 -1) > 0$ implying that these
are stable.

Then the minimal energy supersymmetry breaking brane configuration 
is given by Figure 7. 
If we are moving NS5'-brane to $\pm x^7$ direction, then 
the brane configuration will lead to the Figure 6 of 
\cite{Ahn07-3}
 where the gauge group is $SO(N_c) \times
SU(\widetilde{N}_c')$ or $Sp(N_c) \times SU(\widetilde{N}_c')$,
depending on the movement of NS5'-brane to $+x^7$ direction 
or $-x^7$ direction,
with fundamentals for each gauge group and bifundamentals.

\begin{figure}[ht]
   \epsfxsize=5.0in 
\centerline{\epsffile{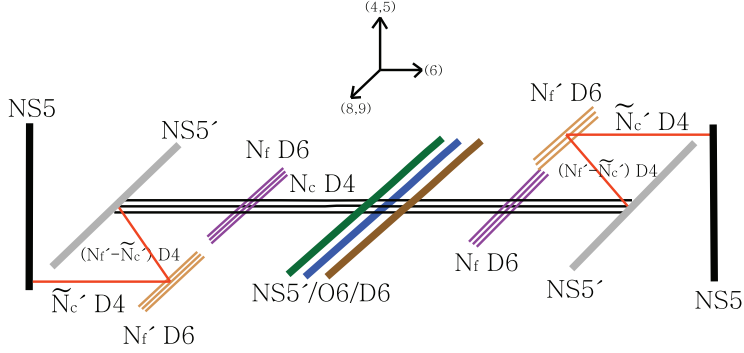}}
   \caption[FIG. \arabic{figure}.]{ 
The nonsupersymmetric minimal energy brane configuration of 
$SU(N_c) \times SU(\widetilde{N}_c'=N_f'+N_c-N_c')$ 
gauge group with fundamentals $Q,
\widetilde{Q}, \hat{Q}, q'$ and $\widetilde{q'}$ for each 
gauge group, the bifundamentals $Y$ and $\widetilde{Y}$ 
and 
an antisymmetric flavor $A$ and a
conjugate
symmetric flavor $\widetilde{S}$ for $SU(N_c)$
and various
gauge singlets.
We consider the massless case of $Q$ and $\widetilde{Q}$.
From the modified Figure 3, there is a misalignment of $N_f'$ flavor
D4-branes.
Some of them are connecting to NS5'-brane and the other to
NS5-brane. In this figure, we put $\theta=0$ and $\theta'=\frac{\pi}{2}$. }
\end{figure}

Now the asymptotic regions of various NS5-branes 
can be determined by reading off the first two terms of quintic 
curve given in subsection 3.2 giving the
$NS5_L$-brane
asymptotic region, next two terms giving 
$NS5'_L$-brane asymptotic region, next two terms
giving $NS5'_M$-brane asymptotic region, next two terms 
giving $NS5'_R$-brane asymptotic region, and 
final two terms giving $NS5_R$-brane asymptotic region
as follows:

1. $v \rightarrow \infty$ limit implies
\bea
w & \rightarrow & 0, \quad y \sim    v^{\widetilde{N}_c'} + \cdots \quad
\mbox{$NS5_{L}$ 
asymptotic region}, \nonu \\
w & \rightarrow  & 0, \quad y \sim    
v^{-\widetilde{N}_c'+2N_f+2N_f'+4} + \cdots \quad
\mbox{$NS5_{R}$ asymptotic region}.   
\nonu
\eea

2.  $w \rightarrow \infty$ limit implies
\bea
v & \rightarrow &   -m', \quad 
y \sim  w^{N_c-\widetilde{N}_c'}
 +\cdots
\quad \mbox{$NS5_{L}'$ asymptotic region}, 
\nonu
\\
v & \rightarrow &   0, \quad 
y \sim  w^{N_f+N_f'+2}
 +\cdots
\quad \mbox{$NS5_{M}'$ asymptotic region}, 
\nonu
\\
v & \rightarrow &  +m', \quad  
y \sim w^{\widetilde{N}_c-N_c+2N_f+2N_f'+4}
+\cdots
\quad \mbox{$NS5_{R}'$ asymptotic region}. 
\nonu
\eea

The two $NS5_{L,R}'$-branes 
are moving in the $\mp v$ direction holding everything 
else fixed instead of moving D6-branes in the $\mp v$ direction.
The corresponding mirrors of D4-branes are moved appropriately.
 Since the extra terms in the harmonic
function contribute to the $g(s)$ as a multiplication factor,   
the statement that $y$ is not equal to zero even if $v=0$,
implies that there exists the instability from a new M5-brane mode at
some point from the transition of SQCD-like theory description 
to M-theory description.

It is natural to ask if we can generalize the procedure for the
product gauge group of two gauge groups, which will contain three
NS5-branes, four NS5-branes or five NS5-branes,
to configurations with more than the product gauge
group of three gauge groups. When there exist three NS5-branes with
two gauge groups, one
can add an extra NS5-brane to the left of left NS5-brane or to the right
of right NS5-brane for the triple gauge groups. 
This extra NS5-brane will be perpendicular to its
closest NS5-brane. One can also add an orientifold 4-plane in this
brane configuration. When there are four NS5-branes with O6-plane with
two gauge groups, 
we add two outer NS5-branes in a ${\bf Z}_2$ symmetric way and each
outer NS5-brane will be perpendicular to its closest NS5-brane for the
triple gauge groups.
When there are five NS5-branes with two gauge groups, 
one adds  two outer NS5-branes in a ${\bf Z}_2$ symmetric way and each
outer NS5-brane will be perpendicular to its closest NS5-brane for the
triple gauge groups.   
For each case, one needs to understand how the magnetic superpotential
arises from its electric theory 
and to analyze the correponding F-term equations. 

There exist different directions concerning on the meta-stable vacua
in different contexts. Some of the relevant works are present in
recent works \cite{GSU}-\cite{ABFK} where some of them use anti D-branes 
and some of them are described in the type IIB theory.
It would be interesting to find out
how similarities and differences between type IIA and IIB theories 
arise. 

\vspace{.7cm}

\centerline{\bf Acknowledgments}

I would like to thank 
D. Shih and K. Landsteiner 
for discussions and Harvard High Energy Theory Group for hospitality
where part of this work was undertaken. 
This work was supported by grant No.
R01-2006-000-10965-0 from the Basic Research Program of the Korea
Science \& Engineering Foundation.

\end{document}